# Observation of second sound in a rapidly varying temperature field in Ge

Albert Beardo[1], Miquel López-Suárez[2,3], Luis Alberto Pérez[2], Lluc Sendra[1], Maria Isabel Alonso[2], Claudio Melis[3], Javier Bafaluy[1], Juan Camacho[1], Luciano Colombo[3], Riccardo Rurali[2], F. X. Alvarez[1], and Juan Sebastián Reparaz[2]*

**Affiliations:**

[1]Departament de Física, Universitat Autònoma de Barcelona, 08193 Bellaterra, Spain.

[2]Institut de Ciència de Materials de Barcelona (ICMAB-CSIC), Campus de la UAB, 08193 Bellaterra, Spain.

[3]Dipartimento di Fisica, Università di Cagliari, Cittadella Universitaria, I-09042 Monserrato (Ca), Italy.

*Corresponding author: jsreparaz@icmab.es

**Abstract:** Second sound is known as the thermal transport regime where heat is carried by temperature waves. Its experimental observation was previously restricted to a small number of materials, usually in rather narrow temperature windows. We show that it is possible to overcome these limitations by driving the system with a rapidly varying temperature field. This effect is demonstrated in bulk natural Ge between 7 kelvin and room temperature by studying the phase lag of the thermal response under a harmonic high frequency external thermal excitation, and addressing the relaxation time and the propagation velocity of the heat waves. These results provide a new route to investigate the potential of wave-like heat transport in almost any material, opening opportunities to control heat through its oscillatory nature.

**One Sentence Summary:** Wave-like heat propagation is shown to emerge as a consequence of a rapidly varying temperature field in Ge, and thus possibly in other materials as well.

**Main Text:** The study of heat transport beyond Fourier's regime has attracted renewed interest in recent years. Great efforts have been performed to unravel the physical properties of thermal waves, as well as the experimental conditions that are necessary for their observation. Applications based on such concepts have been envisioned and discussed extensively already in many recent publications (1–4). The spatio-temporal propagation of the temperature field in the form of waves is known as "second sound," a term that was adopted in analogy to "first sound" (mechanical lattice vibrations). As pointed out in Ref. (5), first and second sound are described by a similar equation where the variables have a different physical meaning, i.e. pressure and temperature, respectively.

The simplest differential equation that best describes heat transport from a mesoscopic perspective is the hyperbolic heat equation (HHE) due to Maxwell, Cattaneo, and Vernotte:

$$\tau_{ss}\frac{\partial^2 T}{\partial t^2} + \frac{\partial T}{\partial t} - \alpha\nabla^2 T = \frac{1}{\rho C_p}S(\vec{r},t) \qquad (1)$$

where $\alpha$ is the thermal diffusivity, $\tau_{ss}$ is the thermal relaxation time, $\rho$ is the mass density, $C_p$ is the specific heat, and $S(\vec{r},t)$ is an external power heat source. In our context, the system is in local-equilibrium and it is well characterized by a local temperature $T$ (see discussion in SM8). The previous equation describes the propagation of a temperature wave with a damping term given by $\partial T/\partial t$ and a propagation velocity $v_{ss} = \sqrt{\alpha/\tau_{ss}}$. The solutions of this equation lead to different heat transport regimes depending on the temporal and spatial length scales under investigation. The key to unlock the different regimes is the magnitude of the first term on the left-hand side of Eq. (1), thermal inertial term, i.e. if $\tau_{ss}$ or $\partial^2 T/\partial t^2$ are sufficiently large, the spatio-temporal distribution of temperature field will exhibit wave-like behavior.

Second sound in solids was first experimentally observed in solid He (6), later in NaF (7), Bi (8), SrTiO$_3$ (9), and most recently in highly oriented pyrolytic graphite (10). Several theoretical works have also recently addressed its occurrence in low dimensional systems (11–13). In all experimental observations of second sound the dominance of momentum conserving phonon scattering (Normal processes) with respect to resistive phonon scattering (Umklapp processes) was found to be the key mechanism leading to its observation. In fact, second sound was observed almost exclusively in the very low temperature regime (T < 5K), with the exception of a recent example (10) at high temperatures (125 K) for samples with low resistive phonon scattering. A condition for the experimental detection of second sound, based on these experimental observations (6–10), was found to be: $\tau_N < \tau_{exp} < \tau_R$, i.e. the typical experimental observation times ($\tau_{exp}$) must be larger than normal phonon scattering times ($\tau_N$) to allow momentum redistribution but smaller than resistive phonon scattering times ($\tau_R$) to avoid decay of the phonon wave-packet into the phonon equilibrium distribution. So far, second sound was experimentally observed exclusively in the limit of weak resistive phonon scattering (6–10). Although early theoretical works (5,14,15) suggested that a rapidly varying temperature field would provide an alternative mechanism to arrive at the evolution described by the HHE, this effect had never been observed experimentally.

We show that it is possible to observe second sound in natural bulk Ge by driving the system out of equilibrium with a rapidly varying temperature field. Our concept is based on taking advantage of the second order time derivative in the HHE, Eq. (1), in a frequency-domain experiment. As the driving frequency increases towards the hundreds of MHz range, the relative weight of this term with respect to the damping term (first order time derivative) increases proportionally to the frequency upon a harmonic excitation, hence making possible the observation of wave-like heat propagation. We show that this approach is robust enough to expose second sound independently, to a certain extent, of the phonon scattering rates of the studied material, as well as of temperature. In fact, although heat transport in Ge is dominated by resistive phonon scattering processes, which partly originate from its large isotopic diversity, we show that it is still possible to observe second sound in the high frequency limit.

Our experiments are based on a frequency-domain optical reflectance pump-and-probe approach based on two lasers with different wavelengths ($\lambda_{pump}$=405 nm, $\lambda_{probe}$=532 nm) focused onto the surface of a Ge sample to a spot size with radius, $R_{spot}$≈5.5 μm. The studied samples are pieces of a substrate of natural Ge. Further details are provided in the Supplementary Materials (SM1). The pump laser (thermal excitation) is modulated between 30 kHz and 200 MHz with a sinusoidal power output waveform, leading to a dynamic modulation of the optical reflectivity of the surface of the sample, which is also well described by a harmonic waveform. A frequency-dependent phase lag gradually develops, defined as a phase difference between the harmonic thermal excitation, $S(\vec{r}, t)$, and the response of the sample, $T(\vec{r}, t)$ (SM2 and SM7), which can be modelled using Eq. (1) in the present experimental conditions. The choice of Ge as candidate for the observation of second sound is not arbitrary, and it is mostly based on the large optical absorption coefficient of this material for the wavelengths used in this experiment. In fact, the optical penetration depth of the pump and probe lasers is $\delta^{405nm}$=15 nm and $\delta^{532nm}$=17 nm, respectively (SM3). These particular conditions make Ge an ideal material for this study, the small penetration depth of both lasers ensure that the measured phase lag is local and, thus, accurately describes the oscillations of the thermal waves (16). We note that wave-like effects are not observed in the presence of a metallic transducer (SM9), since the thermal interface between the transducer and the Ge substrate, as well as the transducer itself, dominate the system response in the frequency range where wave-like effects are expected.

In the present experiments the changes $\Delta R$ of the optical reflectivity upon the pump laser excitation are driven by the lattice temperature variation, hereafter referred to as $\Delta T$, and by the modulation $\Delta n$ of the carrier concentration in the conduction band. The resulting $\Delta R$ at the energy corresponding to the probe wavelength is provided by the first-order expansion $\Delta R = (\partial R/\partial T)\Delta T + (\partial R/\partial n)\Delta n$ (17). Interestingly enough, the contribution provided by the variation of the carrier concentration is expected to dominate the way the reflectivity is affected in experiments, like the present ones, involving pulsed laser sources, i.e. for high electronic excitation densities (18,19). However, in our excitation conditions the electronic contribution to the optical reflectivity can be neglected, hence, the optical reflectivity is dominated by the temperature of the lattice ($\Delta T$) for all excitation frequencies. We demonstrate this by studying the relative magnitude of $(\partial R/\partial T)\Delta T$ and $(\partial R/\partial n)\Delta n$. In particular, we take advantage of the fact that for bulk Ge, $\partial R/\partial T \approx 0$ at $T = 220\ K$ for 532 nm of probe wavelength (see SM3), whereas $\partial R/\partial n$ is expected to be temperature almost independent between 220 K and room temperature. Furthermore, $\partial R/\partial T$ exhibits a sign inversion at this temperature which is reflected in a change of the phase lag by an angle of $\pi$. Figure 1A displays the temperature dependence of $\Delta R/R$, as well as the phase lag for low (30 kHz) and high (100 MHz) modulation frequencies, and for a constant pump power of ≈10 mW. The reflectivity signal exhibits a minimum (20) around ≈220 K, followed by a slow signal recovery at lower temperatures (21). On the other hand, the phase lag is shifted by $\pi$ at ≈220 K (140º → -40º). The optical reflectivity extracted from ellipsometry experiments corresponding to $\Delta T = 1K$ are also shown for relative comparison. A similar behavior is observed independently of the excitation frequency, which resembles the ellipsometry data for which $\Delta n = 0$. In fact, from the reflectivity signal ratio between the minimum observed in Fig. 1A and the value of the reflectivity at room temperature we can estimate an upper boundary for $(1/R)[\partial R/\partial n] \approx -2\times 10^{-29}$ m³, assuming that the residual signal at ≈220 K has purely electronic origin. The inset of Fig. 1A displays calculations of the electronic (solving the electron recombination diffusion equation) and lattice (solving HHE) contributions to the optical reflectivity as a function of frequency (SM3). Whereas the lattice contribution to the optical reflectivity is at least 20-fold larger than the electronic component at the higher frequencies, for lower frequencies this ratio is substantially increased.

Figure 1B displays the experimental phase lag as a function of frequency between 30 kHz and 200 MHz at room temperature. The complex thermal response of the specimen was at first computed numerically within Fourier's model, solving the parabolic approximation to the 3-dimensional (3D) HHE (diffusive case), which is obtained when the first term of Eq. (1) can be neglected. In the range between 30 kHz and 1 MHz, the agreement between Fourier's solution and the experimental data is excellent, although deviations are already observed around 1 MHz. Above 30 MHz the difference between the experimental phase lag and Fourier's predictions is evident. In fact, for the higher frequency range, the experimental data show that the phase lag (absolute value) decreases with increasing frequency. This trend cannot even be qualitatively reproduced by Fourier's model, which predicts that as frequency increases, the phase lag approaches $-\pi/4$ and even lower values (see SM7 and Fig. SM7-3). The full 3D solution of the HHE based on the finite element method was used to fit the experimental data through the entire frequency range, and it is shown in Fig. 1A. A detailed description of the fitting procedure is presented in SM7. We obtained $\tau_{ss}^{exp}$=500 ps and $\alpha^{exp} = 3\times 10^{-5}$ m²/s, thus leading to a propagation velocity $v_{ss}^{exp} = 250$ m/s, all at room temperature. These experimental observations were numerically confirmed through computational experiments (see schematic illustration in Fig. 1C) by non-equilibrium molecular dynamics (NEMD), and are shown in the inset of Fig. 1B (see also SM5 and Fig. SM5-1). Although a quantitative agreement cannot be expected due to differences between the experimental and the computational setups (reduced size of the sample and purely 1D heat flux in NEMD), the NEMD results show a striking similarity with the experimental results, with the phase lag initially decreasing, hitting a minimum, and then recovering. These results are of particular interest since, within the NEMD approach, no assumption is made regarding the heat transport regime. Thus, the numerical experiments are an independent confirmation of the appearance of second sound in the high frequency limit.

The frequency window where second sound is expected can be estimated comparing the thermal penetration depth of the diffusive and wave-like regimes. Figure 1D displays the penetration depth, $\Lambda_{HHE}$, calculated using the exact solution of the HHE (see SM7), as well as the diffusive and wave-like limits, $\Lambda_{diff} = \sqrt{\alpha/(\pi f)}$ and $\Lambda_{ss} = 2\sqrt{\alpha \tau_{ss}}$, respectively. A critical frequency $f_c$ is obtained when $\Lambda_{diff} = \Lambda_{ss}$, thus, providing an estimation of the frequency for which the diffusive and wave-like contributions to heat transport are similar. The temperature dependence of $\tau_{ss}$, $v_{ss}$ and $f_c$ was studied between room temperature and 7K, and it is shown in Fig. 2A (full set in SM2). As temperature decreases, the ratio between the penetration depth of the wave-like and diffusive contributions is $\Lambda_{ss}/\Lambda_{diff} = \sqrt{4\pi f \tau_{ss}(T)}$, which implies that lower temperatures favor the spatial propagation of the thermal waves since larger $\tau_{ss}$ are expected and indeed experimentally observed for lower temperatures. In fact, wave-like effects are already present below $f_c$, as can be observed comparing the experimental data with the corresponding fits using the HHE, to the Fourier predictions as shown in Fig. 2A (22). The onset of wave-like effects is also evidenced by the deviations between $\Lambda_{diff}$ and $\Lambda_{HHE}$ as shown in Fig. 1D. The minimum observed on the phase lag curves as well as the position of the critical frequency $f_c$ relative to the frequency of the minimum, originate from the relation between $\alpha$, $\tau_{ss}$, and $R_{spot}$ (see discussion in SM7). The frequency dependent phase lag in Fig. 2A was fitted at each temperature (23) using the HHE, as described for the room temperature case, and the results for $\tau_{ss}^{exp}$ and $v_{ss}^{exp}$ are shown in Fig. 2B and in Table SM2.

To understand the origin of these observations we have developed a rather simple model (see derivation in SM6) based on the expansion of the perturbed phonon distribution function (24), i.e. the intermediate state assumed by the non-equilibrium distribution before it decays to the equilibrium one by means of dissipative resistive processes, as: $f_\lambda = f_\lambda^{eq} + \vec{\beta}_\lambda \cdot \vec{q} + \vec{\gamma}_\lambda \cdot (\partial \vec{q}/\partial t)$, where $f_\lambda^{eq}$ is the equilibrium phonon distribution function, $\vec{q}$ is the heat flux, $\vec{\beta}_\lambda$ and $\vec{\gamma}_\lambda$ are mode dependent functions to be determined, $t$ is the temporal coordinate, and $\lambda$ denotes each phonon mode. We note that the case with $\vec{\gamma}_\lambda = 0$ leads to the same propagation velocity for second sound as in Ref. ($^5$) (SM6). In our case, however, the presence of rapidly varying temperatures leads to rapidly varying $\vec{q}$, suggesting that the expansion of the perturbed phonon distribution function in terms of $\partial \vec{q}/\partial t$ is a reasonable assumption. The previous *ansatz* for $f$ was then introduced into the linearized Boltzmann transport equation (BTE) to find the solution for $\vec{\beta}_\lambda$ and $\vec{\gamma}_\lambda$. It can be shown that, within this framework, an explicit expression for $\tau_{ss}$, in terms of individual phonon relaxation times ($\tau_\lambda$), can be obtained as (SM6):

$$\tau_{ss} = \frac{\sum_\lambda \hbar \omega_\lambda v_\lambda^2 \tau_\lambda^2 \frac{\partial f_\lambda^{eq}}{\partial T}}{\sum_\lambda \hbar \omega_\lambda v_\lambda^2 \tau_\lambda \frac{\partial f_\lambda^{eq}}{\partial T}} \qquad (2)$$

where $\hbar$ is the Planck constant, $\omega_\lambda$ is the phonon energy, and $v_\lambda$ is the phonon group velocity. We computed $\omega_\lambda$, $\tau_\lambda$, and $v_\lambda$ from the solution of the BTE based on Density Functional Theory (DFT) interatomic force constants (SM4 and Table SM4). We restricted ourselves to the Relaxation Time Approximation (RTA) after verifying that the full iterative BTE picture does not alter the prediction of the theory. Using the RTA has the additional benefit of providing a comparison on equal footing with previous theoretical descriptions (5), and allowing unambiguous definition of the relaxation times (25). In fact, we observe that the corrections to the RTA provided by a full iterative solution of the BTE are very small (within 4% in the thermal conductivity) for Ge at temperatures as low as 50 K (Fig. SM4-1). The resulting values were inserted into Eq. (2), which yielded $\tau_{ss}^{theo}$ and $v_{ss}^{theo} = \sqrt{\alpha^{theo}/\tau_{ss}^{theo}}$ as a function of temperature, as shown in Fig. 2A. The agreement of the predicted values with those obtained from the experiments is remarkable for T>100 K, considering that the values of $\omega_\lambda$, $\tau_\lambda$, and $v_\lambda$ are evaluated within a fully ab initio scheme. We note that the model leading to Eq. (2) is expected to be valid for $|\Delta T| \ll T$, where $|\Delta T|$ is the amplitude of the laser induced thermal oscillations and $T$ is the absolute temperature as set by the cryostat. In our experiments, $|\Delta T|_{max} \approx 10$ K, thus the observed deviations between the theoretical predictions and the measured values at very low temperatures are expected (see SM8 for details).

The spatial dependence of the temperature field in the parabolic (diffusive) and the hyperbolic (wave-like) cases was simulated using finite element methods in the direction perpendicular to the surface of the sample at an arbitrary time. Figure 2C displays the normalized temperature profiles for 15 K, 100 K, and 300 K at the highest experimental excitation frequency of ≈300 MHz (see Fig. SM7-5 for simulations of the temperature field at $f_c$). As expected, the wave-like behavior of the temperature field exhibits a strong temperature dependence. The observed propagation depth, particularly at lower temperatures, is especially interesting if considering the possibility of high frequency modulated thermal interference.

We think that the present approach could open new possibilities for experimental observation of wave-like heat transport in other materials and lead to the development of novel strategies to control heat transport.

**Acknowledgments:** The authors acknowledge Dr. Mariano Campoy-Quiles for fruitful scientific discussions and for a critical reading of the manuscript, and Prof. Keivan Esfarjani for discussions during the development of the BTE model. **Funding:** The authors acknowledge financial support from the Spanish Ministry of Economy, Industry, and Competitiveness through the "Severo Ochoa" Program for Centers of Excellence in R&D (SEV-2015-0496), MAT2017-90024-P (TANGENTS)-EI/FEDER, Grant No. RTI2018-097876-B-C22 (MCIU/AEI/FEDER, UE), and by the Generalitat de Catalunya under grant no. 2017-SGR-1506 and 2017-SGR-00488. We thank the Centro de Supercomputación de Galicia (CESGA) for the use of their computational resources. MLS was funded through the Juan de la Cierva programme. A.B.R., LL.S.M, J.B.B., J.C.C and F.X.A. acknowledge financial support by Spanish Ministerio de Ciencia, Innovación y Universidades. **Author contributions:** This work was conceived and led by J.S.R. The experiments were done by L.A.P, M.I.A. and J.S.R. Non-Equilibrium Molecular Dynamics (NEMD) simulations were conceptualized, executed, and interpreted by M.L.S, C.M., and L.C. Finite elements modelling was done by A.B. and X.A. Ab-initio simulations by R.R. Modelling based on BTE by A.B., L.S., J.C., J.B, X.A. All authors contributed in drafting the manuscript and participated in the scientific discussion. **Competing interests:** Authors declare no competing interests. **Data and materials availability:** Most data are available in the main text or the supplementary materials. Additional data that support the findings of this study are available from the corresponding authors upon reasonable request.

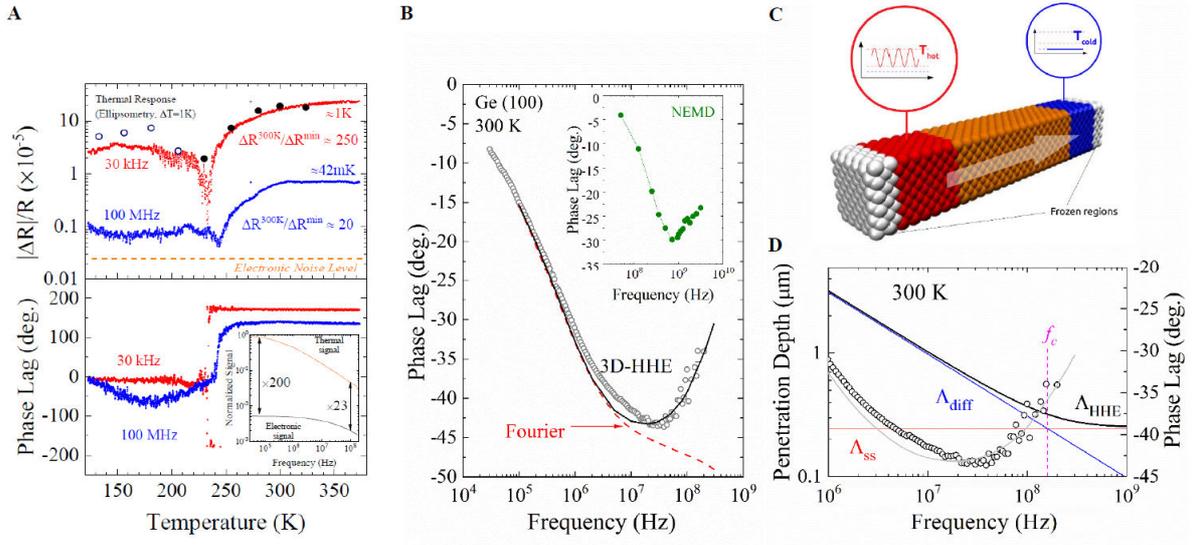

**Figure 1.** (**A**) The upper panel displays the optical reflectivity change as a function of temperature for 30 kHz and 100 MHz. The back dots are the results obtained from ellipsometry measurements to a temperature rise of 1 K extracted from SM3 (full symbols→dR/dT<0, open symbols→dR/dT>0). The lower panel displays the phase lag of the signal with respect to the pump excitation. The inset displays calculations accounting for the thermal and electronic contributions to the reflectivity. (**B**) The experimental phase lag at room temperature between the pump and probe lasers as a function of the pump excitation frequency is shown in black open symbols. The inset displays numerical experiments using NEMD in green full dots. In dashed red line we display the prediction based on Fourier's law. The solutions based on the 3D-HHE are shown in black line with a resulting fitted $\tau_{ss}$ = 500 ps. (**C**) Schematic illustration of the geometry used for the NEMD numerical experiments. The red (blue) regions correspond to the regions connected to the hot (cold) thermostat, while the regions in white are kept frozen. Heat transport and the development of the phase lag is studied in the central, orange region. (**D**) Frequency dependent thermal penetration depth calculated using the solution of the HHE ($\Lambda_{HHE}$), the diffusive case ($\Lambda_{diff}$), and the penetration depth ($\Lambda_{ss}$) obtained in the high frequency limit from Eq. (1). The high frequency experimental phase lag is shown for comparison. The crossover between both curves defines the frequency, $f_c$, where $\Lambda_{diff} = \Lambda_{ss}$.

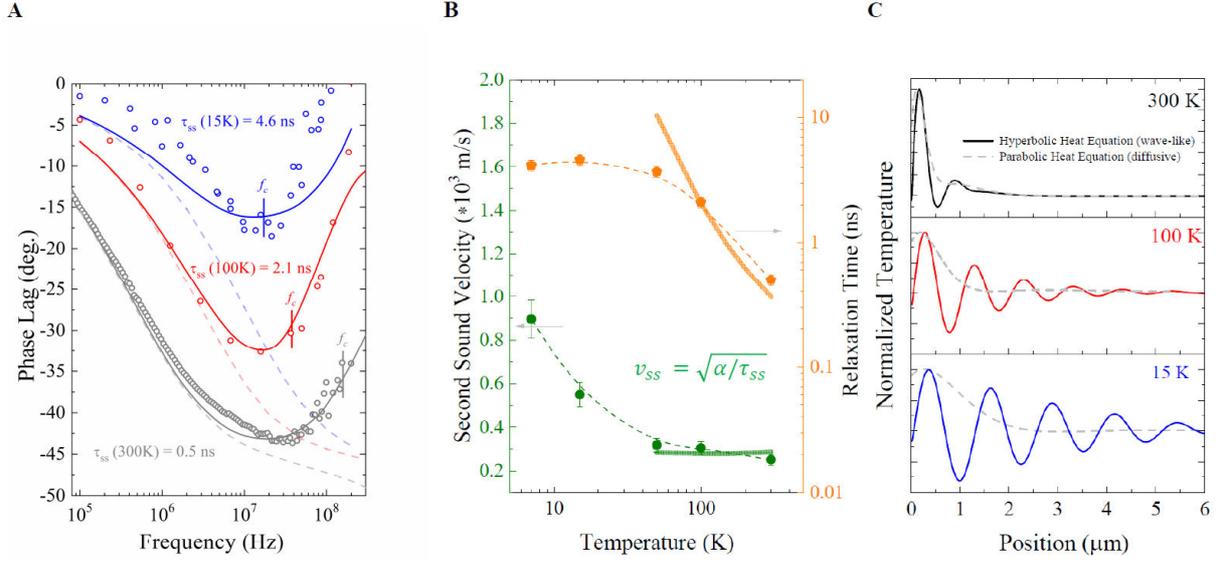

**Figure 2.** (**A**) Phase lag versus frequency for the higher frequency range as a function of temperature. We plot three points at 300 K, 100 K, and 15 K with the corresponding fits to the data point using the 3D-HHE. In dashed lines we display the prediction based on Fourier's law at each temperature. (**B**) The experimental relaxation times ($\tau_{ss}$), as well as the propagation velocity ($v_{ss}$) are shown as a function of temperature. The dashed lines are guides to the eye. The full lines are the predictions based on the expansion of the perturbed phonon distribution function, $f = f_\lambda^{eq} + \vec{\beta}_\lambda \cdot \vec{q} + \vec{\gamma}_\lambda \cdot (\partial \vec{q}/\partial t)$, combined with DFT simulations as described in SM6 and SM4, respectively. (**C**) Finite element simulations of the spatial distribution of the temperature field at a function of temperature in the direction perpendicular to the surface of the sample at the highest excitation frequency of ≈300 MHz for an arbitrary time. The parabolic and hyperbolic solutions are shown in dashed and full lines, respectively.

# Supplementary Materials:

## "Observation of second sound in a rapidly varying temperature field in Ge"

**Materials and Methods**

<u>SM1</u> - Samples description & processing

All samples were pieces cleaved from a 2-inch diameter nominally undoped Ge wafer with (100) crystallographic orientation, high resistivity (> 40 $\Omega$ cm), and etch pit density (EPD) < 3000 /cm². The wafer was purchased from International Wafer Service Inc. (USA). We have studied three different types of samples:

(i) <u>Bare Ge:</u> the sample was rinsed in acetone and dried under $N_2$ flux. A layer of native oxide with a thickness of ≈3 nm was observed at the surface of the sample, as determined from spectroscopic ellipsometry measurements. We have also studied a similar sample without the native oxide layer, which was removed using the procedure described in (iii). However, no influence of the native oxide in the phase lag response was observed in the absence of Au transducer.

(ii) <u>Ge + native oxide + 60 nm of Au:</u> A 60 nm thick Au transducer was evaporated onto the surface of a Ge piece similar to (i). The evaporation chamber was purged using highly pure $N_2$ gas, the base pressure was ≈1x10$^{-7}$ mbar, and the Au deposition rate was set to 0.6 Å/s. The thickness of the Au transducer was measured using atomic force microscopy.

(iii) <u>Ge + 60 nm of Au:</u> this sample is similar to (ii), however, the native oxide layer was stripped by dipping in diluted HF (10% in $H_2O$) for 1 minute. Immediately after, the sample was inserted into the evaporation chamber to prevent $GeO_x$ formation. Au evaporation was simultaneously performed on both samples, (ii) and (iii), to ensure that the Au transducer is similar in both.

## SM2 - Experimental Methods

We developed a low noise custom-built frequency-domain thermoreflectance (FDTR) set-up to measure the thermal response of the samples. Figure **SM2-1** displays a sketch of the experimental arrangement. FDTR is a contactless technique which is based on probing the time-dependent reflectivity of a specimen upon a modulated thermal excitation. The thermal excitation was provided by a pump laser diode with a wavelength of 405 nm purchased from Omicron-Laserage, model A350. This laser module was temperature-stabilized with high-speed analog modulation, maximum output power of 300 mW, and bandwidth of 200 MHz. The probe laser used was a continuous wave (CW) laser from Cobolt (08-01 Series) of 532 nm wavelength and with a maximum output power of 100 mW. The probe laser was optimized (factory settings) for interferometry purposes, thus, providing low noise output and a coherence length > 2 m. Both lasers were coupled into Faraday insulators to prevent back reflections into the cavities of the lasers. A quarter waveplate (one for each laser) was used to homogenize the polarization of the lasers, thus, avoiding any preferential direction. The output power was controlled with neutral density filters to ≈50 µW (CW) for the probe, and ≈20 mW (RMS) for the pump laser. Both lasers were coupled to the same optical path using beam splitters and dichroic mirrors as shown in Figure **SM2-1**. A 30 mm achromatic lens doublet purchased from Thorlabs was used to focus both Gaussian beams onto the same spot, whose size we have measured using the knife-edge method to a $1/e^2$ radius of ≈5.5 µm. As the pump laser (high power) is modulated with a harmonic function generator, the optical reflectivity of the specimen is modulated with a similar waveform, in our case a sinusoidal profile. In other words, the reflected CW probe laser intensity is proportional to the time-dependent optical reflectivity. The operational principle of this method can be easily understood through the following simple expressions:

$$I_{532} = constant, \; I_{405} = I_0[1 + cos\,(\omega t)]$$

$$R_{532} = R_0 + \frac{\partial R}{\partial T}\Delta T \; \rightarrow \; \Delta T = \Delta T_0[1 + cos\,(\omega t + \varphi)]$$

$$R_{532} = R_0 + \frac{\partial R}{\partial T}\Delta T_0[1 + cos\,(\omega t + \varphi)] = \left(R_0 + \frac{\partial R}{\partial T}\Delta T_0\right) + \frac{\partial R}{\partial T}cos\,(\omega t + \varphi)$$

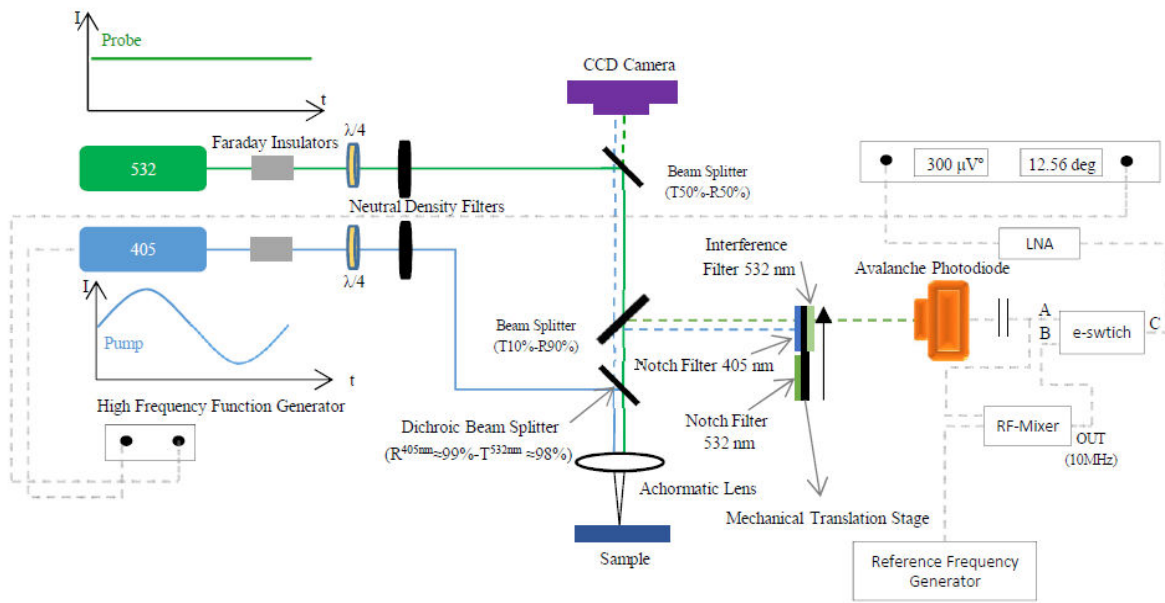
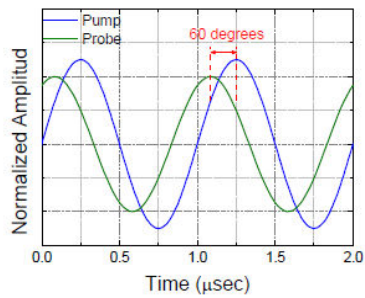

**Figure SM2-1.**
Frequency-domain thermoreflectance setup used to conduct the experiments. The simulation of a typical arbitrary response is shown, as well as the main operational principle based on the modulation of the optical reflectivity arising from a local temperature increase. A detailed list of the specific used components is provided in the setup description section.

where *I* is the power of the pump (405 nm) and probe (532 nm) lasers, ω is the angular frequency (ω=2π*f*, where *f* is the frequency in Hz), *t* is the time coordinate, $R_{532}$ is the optical reflectivity at the probe wavelength, $\Delta T$ is the temperature rise caused by the pump laser, and $\varphi$ is the phase lag between the pump and the probe harmonic signals. We note that $\varphi(\omega)$ is the quantity which can be related to the thermal properties of the studied sample. Therefore, this is the quantity which is measured in the present experiments.

The reflected laser light at the surface of the sample was split into two main components, marked with dashed lines in Figure **SM2-1**. A small fraction of reflected power was focused onto a charge-coupled device (CCD) camera to monitor the shape of the laser spots, as well as the overlap between the pump and probe beams. We note that the overlap was optimized using a piezoelectric-driven mirror within the optical path of the pump laser, in order to maximize the modulated signal arising from the probe laser. A comparatively larger portion of the reflected light was sent to an avalanche photodiode detector purchased from Thorlabs (APD430A2). Two notch filters (Thorlabs) where inserted into the optical path with the purpose of individually blocking the laser components. First, the probe component was blocked and the phase of the pump laser was measured using a high frequency lock-in amplifier (Stanford Instruments SR844). After this phase calibration step was performed, the notch filters mount was mechanically displaced to block the pump laser component, thus, allowing us to measure the harmonic signal arising from the probe laser. Note that only light at 532 nm is measured by the detector due to the presence of an interference filter. In consequence, the latter measured phase lag is that of the induced thermal wave due to the pump laser excitation. This procedure results quite accurate in almost all the frequency range; however, as frequency increases (*f* > 50 Mhz), electronic noise such as, e.g. coherent pickup, increases substantially. In order to overcome this problem, we have used two different approaches: (i) a low noise amplifier (FEMTO) with a 200 MHz broadband was used to amplify the signal well over the coherent pickup limit (typically 1 mV at high frequencies), and (ii) electrical heterodyne mixing was used to reduce the measurement window to a range where coherent noise is almost not present. For this purpose, we used a frequency mixer purchased from Mini-Circuits (ZAD-3+). The measured signal was mixed with a variable high frequency reference to produce a resulting signal at 10 MHz, which is well below the limit for coherent noise pick up.

This approach differs from other methods based on optical heterodyning since we profit from the outstanding performance of electronic mixers to achieve the frequency down conversion.

All measurements were performed with the samples in vacuum conditions at a base pressure of $\approx 10^{-5}$ mbar. Variable temperature measurements were carried out using a He gas-flow cryostat from CryoVac between 7 and 300 K. Figure **SM2-2** displays the full set of temperature measurements as well as the data modelling using the hyperbolic heat equation (see section SM7). The purely diffusive solution is also included for comparison, as well at the value of the critical frequency, $f_c$, for each measured temperature.

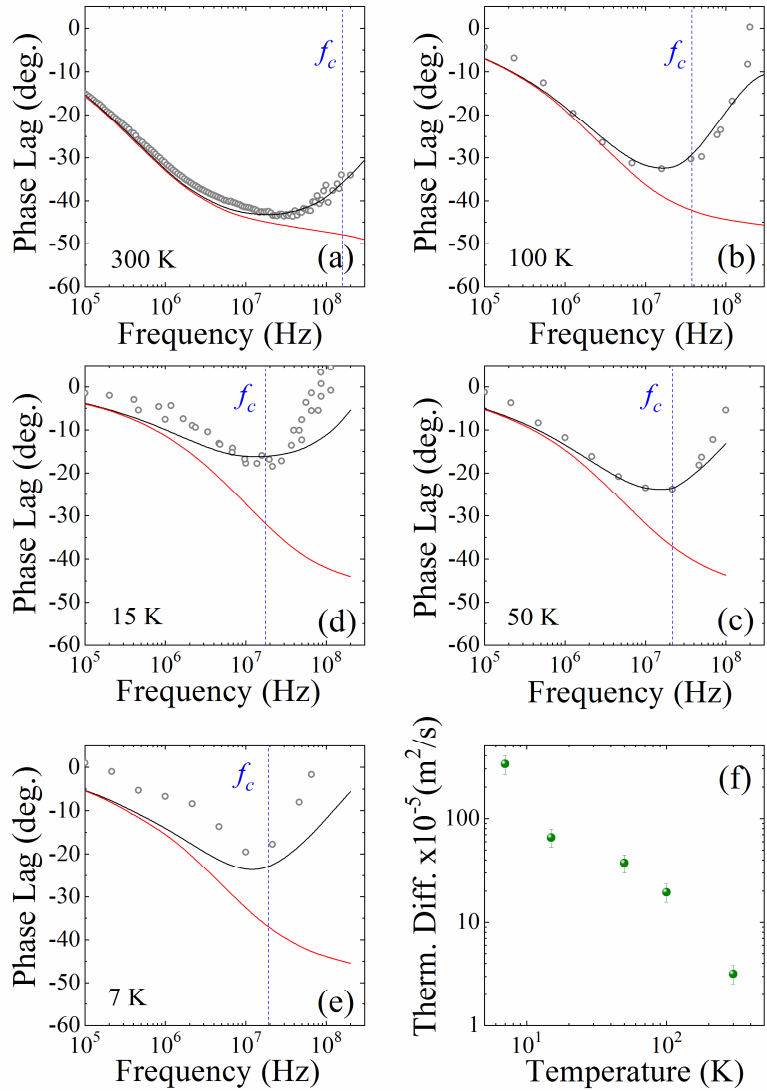

**Figure SM2-2**

(a-e) Phase lag as a function vs. frequency for different temperatures. The higher frequency region was measured in "direct" configuration, i.e. directly measuring the output voltage from the avalanche photodiode, as well as using electrical-heterodyne mixing. Similar results are obtained from both experimental methods. The black lines correspond to fits to the data point using the hyperbolic heat equation, where the relaxation time and the thermal diffusivity have been fitted simultaneously, minimizing the error of the fitting procedure. The red curves correspond to the diffusive solution, as given by the prediction based on Fourier´s law, using the thermal diffusivity as obtained from the hyperbolic fits. (f) The thermal diffusivity as a function of temperature extracted from the fits to the data using the hyperbolic heat equation.

| Temperature [K] | $\tau_{ss}$ (1D-HHE) [s] ×10⁻⁹ | $\tau_{ss}$ (3D-HHE) [s] ×10⁻⁹ | α [m²/s] ×10⁻⁵ | $v_{ss}$ (3D-HHE) [m/s] | $f_c = 1/(4\pi\tau_{ss})$ [Hz] ×10⁶ |
|---|---|---|---|---|---|
| 7 | 6.5 | 4.15 | 333 | 896 | 19.2 |
| 15 | 8.5 | 4.6 | 65.8 | 378 | 17.3 |
| 50 | 4.9 | 3.7 | 37.5 | 318 | 21.5 |
| 100 | 2.1 | 2.1 | 19.5 | 305 | 37.9 |
| 300 | 0.5 | 0.5 | 3.17 | 252 | 159 |

**Table SM2.**

Relaxation times ($\tau_{ss}$), thermal diffusivity (α), and second sound velocity ($v_{ss}$) as a function of temperature, as obtained from fitting the experimental phase lag vs. frequency data (see Figure 1 of the main text), using the 3D hyperbolic heat equation (finite element modelling). We have also fitted the data using the high frequency 1D limit of the hyperbolic heat equation (see analytical derivation in SM7). At high temperatures, the same results are obtained using both approaches. At temperatures below 100 K, the 1D-HHE is not appropriate to fit $\tau_{ss}$ because the system response is 3D even at the highest experimentally available frequencies. Nevertheless, we note that the 1D limit is simpler mathematically and computationally, thus, providing a good compromise solution for cases where finite-elements 3D modelling is not easily available.

The thermal conductivity of the Ge sample was measured independently using the 3 omega method and it is shown in Figure **SM2-3**. We have obtained a thermal conductivity of 51 Wm⁻¹K⁻¹ in good agreement with the values we obtained directly from fitting of the phase lag curves in the frequency domain experiments for $f$ < 1 MHz. Note that for low frequencies and at room temperature, no wave-like effects are observed in the FDTR phase lag data. Furthermore, the value obtained from the 3 omega measurements is not expected to be sensitive to wave-like effects since the technique operates at low frequencies ($f$ < 1 kHz). Excellent agreement was found between the thermal conductivity as obtained from the 3 omega measurements and from fitting the FDTR phase lag data at low frequencies.

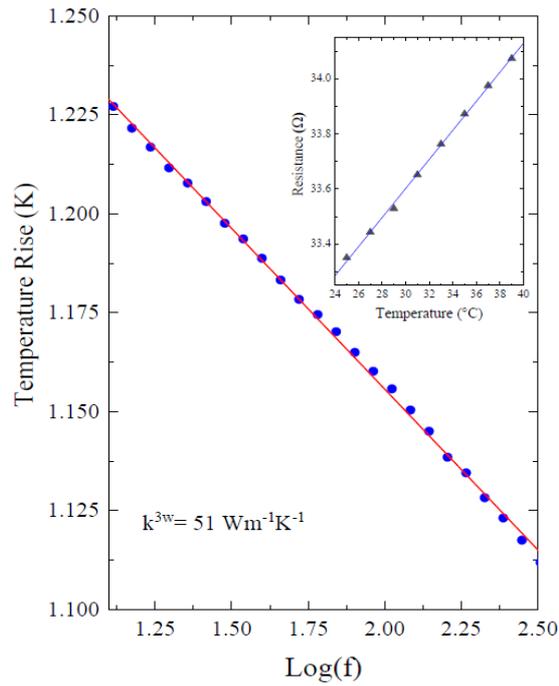

**Figure SM2-3**

Amplitude of the temperature oscillations in the metallic resistor as a function of the heater frequency in logarithmic scale. The inset displays the temperature dependent resistance of the transducer, which is used as thermal sensor. The thermal conductivity of the Ge sample was $\kappa^{3\omega}=51 \, \text{Wm}^{-1}\text{K}^{-1}$.

**SM3** - Thermoreflectance coefficient, optical penetration depth, and hot electrons in Ge

    A. **Thermoreflectance coefficient & optical penetration depth**

Thermoreflectance measurements were performed without the use of a metallic transducer. We note that in typical thermoreflectance experiments for characterization of thermal properties (both in frequency- and time-domain), a metallic transducer is used to ensure a small optical penetration depth, as well as to increase the temperature coefficient of reflectivity and thus the sensitivity of the method. This is typically accomplished by tuning the probe laser wavelength to a plasmonic resonance of the metallic transducer. For 532 nm, the usual choice is Au. Interestingly, Ge has a strong absorption around this wavelength due to the $E_1$ and $E_1+\Delta_1$ interband transitions and as a consequence, measurements without transducer are suitable.

The temperature dependence of the optical quantities was calculated from detailed spectroscopic ellipsometry measurements reported by Emminger et al. (26). Figure **SM3-1** displays the optical reflectivity of Ge as a function of temperature at the wavelength of 532 nm, as well as the temperature coefficient of reflectivity, $C=(1/R)dR/dT$. Since 532 nm is between $E_1$ and $E_1+\Delta_1$, the reflectivity shows a maximum around 215 K due to the combined temperature dependence of the two interband transitions. Therefore, at this temperature, C approaches zero and there is a change of sign from positive at lower T to negative values at higher T. In fact, we have observed a minimum close to the noise level of our detection system in the reflectance curves, in the vicinity of 215 K. Except for this, the magnitude of C is adequate to give a properly measurable signal at most temperatures. It would be problematic again at very low temperatures approaching 0 K where the C tends to zero. The optical penetration depths at selected temperatures are listed in Table **SM3**. The penetration depth is defined from the Beer-Lambert law stating the exponential attenuation of the light intensity as $I = I_0\exp(-z/\delta)$, where $I_0$ is the incident laser power, $1/\delta$ is the absorption coefficient, $\delta$ the penetration depth, and z the spatial coordinate. We note that for both laser wavelengths, pump and probe, the penetration depth is rather small, which allows to consider the thermal excitation and detection virtually as local quantities.

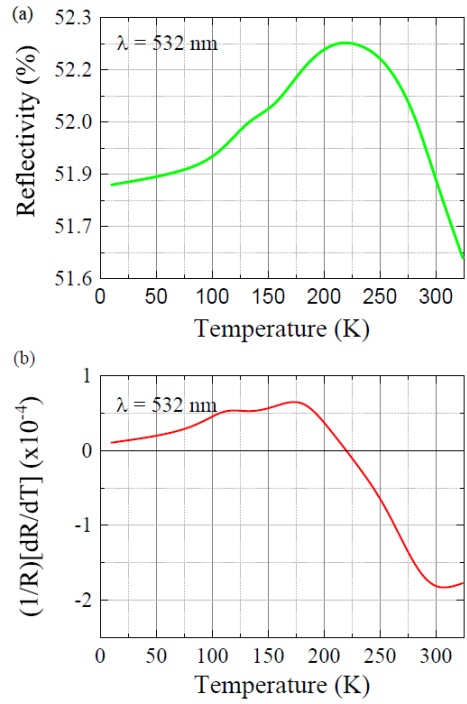

**Figure SM3-1**

Temperature dependence of, (a) the optical reflectivity, and (b) the thermoreflectance coefficient at 532 nm for a Ge (100) substrate, under normal incidence conditions.

| Temperature (K) | $\delta^{405\,nm}$ (nm) | $\delta^{532\,nm}$ (nm) |
|---|---|---|
| 7 | 15.3 | 19.3 |
| 15 | 15.3 | 19.2 |
| 50 | 15.3 | 19.1 |
| 100 | 15.2 | 19.0 |
| 300 | 14.7 | 17.2 |

**Table SM3.**

Optical penetration depth ($\delta$) for the pump (405 nm) and probe (532 nm) laser wavelengths as a function of temperatures, as determined by spectroscopic ellispsometry measurements.

## B. Relaxation of "hot electrons/holes" and heat generation process

It is also worth addressing the heat generation process in Ge which, in the present experiments, takes place through optical excitation of the electronic system. Upon optical excitation with the pump laser at E=3.06 eV vertical transitions are induced from the valence to the conduction band. We recall that the momentum carried by photons is much smaller than the typical momentum of electrons in semiconductors and, thus, the optical excitation process must occur with $|\Delta \mathbf{k}| = 0$. In Ge, the principal "optical" bandgaps ($|\Delta \mathbf{k}| = 0$) in high symmetry points are as follows:

| Optical Transition | $\Gamma_{25'} \to \Gamma_2$ | $\Gamma_{25'} \to \Gamma_{15}$ | $L_{3'} \to L_1$ | $L_{3'} \to L_3$ | $X_4 \to X_1$ |
|---|---|---|---|---|---|
| Energy Gap (eV) | 0.8 | 3.4 | 2.0 | 5.4 | 4.4 |
| Absorption at 3.06 eV | Low | No | High | No | No |

Thus, only electron-hole pairs with $\Delta E \leq 3.06$ eV can be excited, with local densities in $k$-space which depend on the joint density of states between the conduction and valence bands. In fact, for E=3.06 eV excitation, the joint density of states is maximum for states at the $L$ point (see e.g. Figure 1 of Ref. 27). Although electron-hole pairs can also be excited at the $\Gamma$ point, the joint density of excited states at this point is much lower as compared to the $L$ point (see section S3 of Ref. 28), which is also evidenced by the onset of the absorption spectra in Ge at E≈2 eV (29).

After the initial optical excitation process with E=3.06 eV, the "hot electrons" at the $L$ point are scattered to the conduction band minimum, which is located at the same $L$ point. The mechanism which leads to the relaxation of the initially excited electrons is electron-electron scattering, with typical time constants $\tau < 100$ fs (30). On the other hand, the comparatively low amount of electrons excited at the $\Gamma$ point relax towards the $L$ minimum through inter-valley scattering processes in a few hundreds of femtoseconds (31–34). The two previous processes, i.e. (i) initial scattering to local minima and (ii) inter-valley scattering towards the $L$ point, involve the emission of optical and acoustic phonons, hence leading to the production of heat. Note that substantial amount of energy, (3.06 eV – 0.66 eV) = 2.4 eV per absorbed photon, is transferred to the lattice before the electrons reach local equilibrium at the $L$ point. After they reach the minimum of the conduction band at the $L$ point, electrons are already "thermalized" in the sense that they do not produce the emission of thermal phonons. Their relative energy is reduced to E=0.66 eV, compared to the initial E=3.04 eV. Eventually, after several tens of ps to hundreds of µs, depending on the temperature of the lattice, these electrons relax, e.g. through the emission of photons and the absorption (anti-Stokes) or emission (Stokes) of a phonon (required for momentum conservation), Auger processes, defects, etc. We remark that the later recombination processes do not produce heat, since the electrons involved are already "thermalized". These electrons populate the lowest energy states in the conduction band minimum at the $L$ point. Hence, the probability of electron-phonon scattering events, which is the main heat generation mechanism, largely decreases since no lower energy electronic states are available. This implies that almost not heat is created in the electron-hole final recombination process, independently on its radiative or non-radiative nature.

Using the previous values of the relaxation time of the excited electrons, we can estimate an upper limit for the propagation length of the "hot electrons", e.g. using the maximum drift velocity in Ge which at 300 K is $v$=6.5x10$^4$ m/s (35). Hence, the maximum estimated diffusion length for the "hot electrons" $is \Delta x = (6.5\text{x}10^4 \text{ m/s})(100\text{x}10^{-15}\text{s}) = 6.5$ nm. On the other hand, using the observations of Ref. (27) the minimum propagation velocity can be estimated as v=10$^4$ m/s, which results from the size of the studied Ge nanocrystals (11 nm) and the recombination time (1.1 ps) of the electrons at the $L$ point, which is dominated by surface recombination. We obtained a minimum diffusion length for the "hot electrons" of 2 nm. Analogous arguments leading to similar results are also valid for holes (36,37). The estimated propagation length of the "hot electrons/holes", 2 nm to 6.5 nm, can slightly alter the thermal penetration depth as estimated from the optical penetration depth of δ≈15 nm defined as 1/e, (δ≈30 nm as 1/e$^2$). Nevertheless, the results of Figure **SM7-1** show that the effects on the phase lag are almost negligible, and can be of at most 2 degrees.

### C. Electron diffusion recombination equation

In this subsection we show the details of the numerical calculations we have done in order to estimate quantitatively the electronic contribution to the photoreflectance signal. We modeled the "free-electron" density temporal evolution at room temperature using the electron diffusion-recombination equation as described elsewhere (17):

$$\frac{\partial N}{\partial t} = D\nabla^2 N - \frac{1}{\tau_r}(N) - g_2(N)^2 - g_3(N)^3 + S(\mathbf{r})(1 + \sin(2\pi ft))$$

where D is the electron diffusivity, $\tau_r$ is the linear electron recombination lifetime, $g_2$ is the quadratic recombination coefficient (usually associated to band-to-band radiative recombination), and $g_3$ is Auger (cubic) recombination coefficient. Finally, $S(\mathbf{r})$ is the excited electron density introduced in the sample, which has the shape of the pump laser weighted by the amount of absorbed photons per unit time at a given laser power.

In order to provide a conservative estimation of the electron density obtained in experiments, we neglected the non-linear terms in the electron diffusion equation, which in fact amplify the reduction of the amount of excited electrons at high densities. In fact, we have observed that our experimental measurements are independent of the excitation laser power at room temperature, suggesting that the non-linear terms of the electronic diffusion equation are not relevant to compute the total reflectance signal. We assume an electron diffusivity D=25cm$^2$/s and a linear recombination time $\tau_r$=1μs (38), which are conservative values at the average excited densities consistent with our experimental conditions.

In the inset of Figure 1A we show the predicted amplitude of the electron contribution $(\partial R/\partial N)\Delta N$ and the temperature contribution $(\partial R/\partial T)\Delta T$ according to the 3D electron diffusion equation and to the 3D Maxwell-Cattaneo equations, respectively, at a given laser power (i.e. the energy introduced in both simulations is equivalent). Both $\Delta N$ and $\Delta T$ are obtained by convolution with the superficial laser probe beam. The thermoreflectance coefficient at room temperature is $\left(\frac{1}{T}\right)[\partial R/\partial T]$=-1.8x10$^{-4}$K$^{-1}$ as shown in SM3A. Moreover, the electronic reflectance coefficient is expected to be temperature independent and is solely determined by the probe wavelength used, and can be estimated from experiments by varying the excitation power or the frequency at 220 K, where the thermal contribution is suppressed. We obtained $(1/R)[\partial R/\partial n]$=-2x10$^{-29}$m$^3$. Thus, the resulting ratio between the thermal and electronic contributions are in good agreement with the measured signals at 220 K (isolated electronic signal) and at 300 K (electronic and thermal signals combined), as displayed in Fig.1A.

As indicated above, we used a conservative value of the electron diffusivity. By increasing the diffusivity, the electron density in the surface is further reduced along with its contribution to the reflectance signal. In contrast, the electron reflectance signal is independent of the recombination time $\tau_r$. This implies that electron diffusion is the key mechanism reducing the electron density at the observation region, i.e. the surface of the specimens, which cause the corresponding contribution to reflectance to be negligible. Therefore, the higher electron diffusivity with respect to thermal diffusivity in Ge is the main reason why the experimental signal is thermally dominated even at the highest frequencies considered here, which are significantly larger than the electron recombination rates.

**SM4** – Ab initio calculations of the thermal conductivity of Ge

(Note: the equation numbers refer to each SM individual sub-section)

We calculated the second and third order interatomic force constants (IFCs) within density-functional theory (DFT), using the VASP code (39) with the local density approximation (LDA) for the exchange-correlation energy functional and a plane wave cutoff of 174 eV with the projector augmented-wave method. (40,41) The harmonic IFCs were calculated from finite differences in a 5×5×5 supercell, while we used a 4×4×4 supercell for the anharmonic ones, limiting the interactions to fourth neighbors. The inequivalent displacements needed to obtain the IFCs were obtained with the phonopy (42) and thirdorder.py (25) codes. We iterated the electronic self-consistency loop until changes in the total energy and eigenvalues were lower than $10^{-9}$ eV.

Once the IFCs are obtained, we solve iteratively the linearized Boltzmann Transport Equation (BTE) for phonons with the SHENGBTE code (43) on a 24×24×24, **q**-point grid, obtaining the lattice thermal conductivity as,

$$\kappa_l^{\alpha\beta} = \frac{1}{k_B T^2 \Omega N} \sum_\lambda f_0(f_0+1)(\hbar\omega_\lambda)^2 v_{\alpha,\lambda} F_{\beta,\lambda} \qquad (1)$$

where α and β are the three coordinate directions $x$, $y$, and $z$; and $k_B$, $T$, $\Omega$ and $N$ are the Boltzmann constant, the temperature, the volume of the unit cell and the number of **q**-points, respectively. The sum runs over all the phonon modes λ, which have wave vector **q** and branch ν. $f_0$ is the equilibrium Bose-Einstein distribution function, ℏ is the reduced Planck constant, and $\omega_\lambda$ and $v_{\alpha,\lambda}$ are the phonon frequency and phonon group velocity, respectively. $F_{\beta,\lambda}$ is initially taken to be equal to $\tau_\lambda v_{\beta,\lambda}$, where $\tau_\lambda$ is the lifetime of the phonon mode λ within the relaxation time approximation (RTA). Starting from this initial guess, the solution is then obtained iteratively and $F_{\beta,\lambda}$ takes the general form $\tau_\lambda(v_{\beta,\lambda} + \Delta_{\beta,\lambda})$ Scattering from isotopic disorder, besides anharmonic three-phonon scattering, is also included considering the natural isotopic distributions of Ge through the model due to Tamura (43). Within this computational framework we obtain a value for the room temperature thermal conductivity of 51 Wm$^{-1}$K$^{-1}$, in excellent agreement with the experimental values obtained in this work and previously reported. The room-temperature RTA value of the thermal conductivity is 50 Wm$^{-1}$K$^{-1}$, indicating that Normal processes do not play an important role. Similar (negligible) discrepancies are found throughout all the temperature range considered in the *ab initio* calculations

Figure **SM4-1** displays a*b initio* thermal conductivity calculations as a function of temperature for Ge as obtained from the full iterative solution of the BTE (dashed line) and within RTA (continuous line).

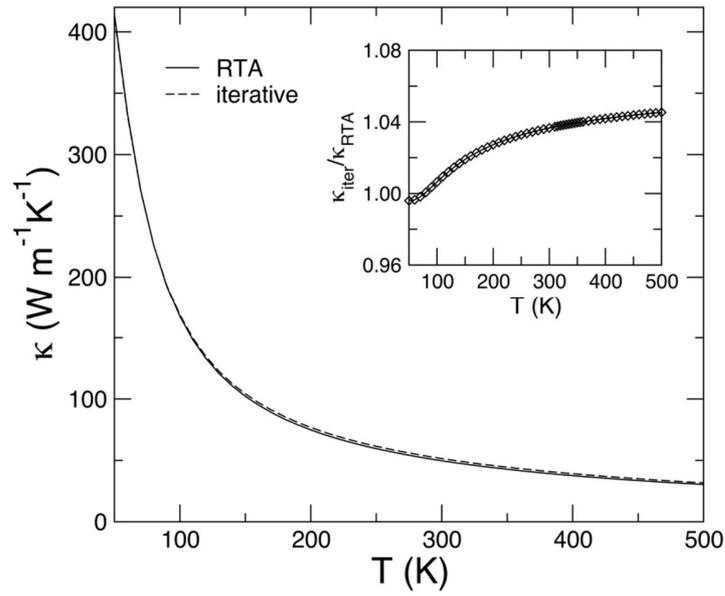

**Figure SM4-1**
*Ab initio* thermal conductivity as a function of temperature of Ge as obtained from the full iterative solution of the BTE (dashed line) and within RTA (continuous line). Note that the temperature axis starts at 50 K. The inset displays the ratio between the two solutions, showing that the corrections to the RTA provided by the iterative solution are negligible throughout all the temperature range considered.

| Temperature [K] | $\tau_{ss}$ [s] ×10⁻⁹ | $\alpha$ [m²/s] ×10⁻⁵ | $v_{ss}$ [m/s] | $f_c$ [Hz] ×10⁶ |
|---|---|---|---|---|
| 50 | 10.4 | 83.4 | 283 | 7.65 |
| 100 | 2.05 | 15.9 | 279 | 38.8 |
| 150 | 0.965 | 7.54 | 279 | 82.5 |
| 200 | 0.622 | 4.94 | 282 | 128 |
| 250 | 0.460 | 3.70 | 284 | 173 |
| 300 | 0.366 | 3.04 | 288 | 217 |
| 350 | 0.304 | 2.53 | 288 | 262 |
| 400 | 0.261 | 2.20 | 291 | 305 |
| 450 | 0.228 | 1.97 | 294 | 349 |
| 500 | 0.203 | 1.78 | 296 | 392 |

**Table SM4.**

Relaxation times ($\tau_{ss}$), thermal diffusivity ($\alpha$), and second sound velocity ($v_{ss}$) as a function of temperature, as obtained from DFT calculations using the corresponding expressions provided in SM4 and SM6.

SM5 - Non-equilibrium molecular dynamics calculations

(Note: the equation numbers refer to each SM individual sub-section)

We carried out computational experiments within non-equilibrium molecular dynamics (NEMD), using a 5×5×145 supercell of the 8-atom cubic Ge conventional cell. The interatomic interactions were described by a bond-order potential of the Tersoff type (44). All the NEMD simulations were performed using the LAMMPS code (45). The equations of motion were integrated with a time step of 1 fs and temperature control was obtained by Nosé-Hoover thermostatting, while equations of motions have been integrated by the velocity-Verlet algorithm. Periodic boundary conditions were applied in all directions. In order to constrain heat to flow through the sample and not through the virtual interface created by the periodic boundary conditions along the $z$ axis, the atoms belonging to the region $0 < z < 12$ Å were kept frozen.

Taking into account the Tersoff parametrization considered in the present work, the size of the frozen region large enough to prevent cross-talking between the hot and cold slabs through the interface. The Ge sample was initially kept for 2 ns under a thermal bias along the $z$ direction of $\Delta T$=-90K, keeping an average temperature of 300 K and thus producing a net heat flow. We neglected the explicit light-matter interaction underlying the heating mechanism in the experimental realization, and rather drive the oscillatory heat flux with a given frequency, $f_p$, by varying the temperature of the hot thermostat as $T_{hot} = T_0 + \Delta T \, Sin\,(f_p t)$. By setting the amplitude of the oscillation to $\Delta T$ we assure that the heat flux will flow from the hot thermostat to the cold one, since $T_{hot}(t) > T_{cold}(t), \forall t$.

The dynamical response of the system was simulated for different frequencies in the 10 MHz–10 GHz range for at least 20 cycles. The actual value of the inward heat flux, $q(t)$, can be obtained by calculating the time derivative of the work performed by the thermostat. We also mapped the temperature along the sample by averaging every 1 ps, obtaining well resolved $T(z,t)$ profiles in the considered frequency range. Figure **SM5-1** displays the $W_{AC}$ and $T(0,t)$ time series corresponding to a pumping frequency of $f_p = 1$ GHz, showing the phase lag between the two oscillations; $W_{AC}$ is the oscillating component of $W$, obtained subtracting from the latter its continuous component $W_{DC}$.

The phase lag between temperature and work was computed as

$$\varphi_{WT} = \cos^{-1}\left(\frac{(T(0,t) \cdot W_{AC}(t))}{|T(0,t_i) \cdot W_{AC}(t_i)|}\right) \qquad (20)$$

Regarding the relation between the heat flux-T phase lag and work-T phase lag, the functional form of the temperature and work time-series can be approximated by trigonometric functions as, for instance, $T(0,t) = T_0 + \Delta T \cdot \sin(\omega t)$ and $W(t) = W_0 \cdot \sin(\omega t + \varphi_\omega)$, where we have written explicitly the dephasing as $\varphi_\omega$. Therefore, the heat flux (taken to be the time derivative of the work performed by the thermostat) can be approximated to $Q(t) = \partial W/\partial t = Q_0 \cos(\omega t +$ $\varphi_\omega$ ), that can be also written by using of a sinusoidal function as $Q(t) = Q_0 \sin(\omega t + \varphi_\omega + \pi/2)$. By comparing the functional forms of the heat flux and the temperature it is clear that the dephasing between them equals $\varphi + \pi/2$. Finally, after computing numerically the phase lag between the computed temperature and work, i.e. $\varphi_\omega$, we can obtain the phase lag between temperature and heat flux as $\varphi_q = \varphi_\omega + \pi/2$.

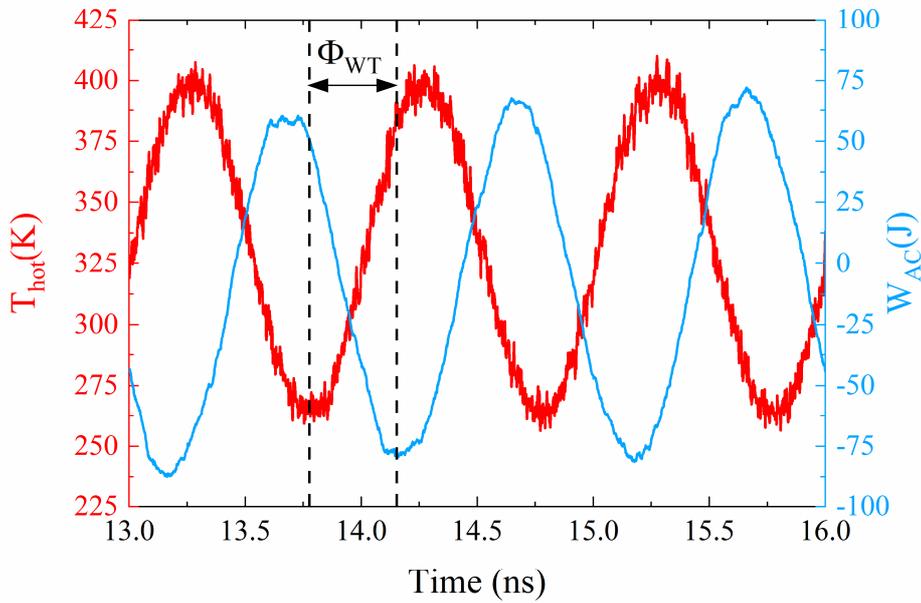

**Figure SM5-1**
Time dependence of the temperature at *z=0*, *T(0,t)*, and of the work of the hot thermostat, W$_{AC}$(J).

**SM6 - Derivation of the Maxwell and Cattaneo heat flux equation from the BTE**

*This section will be updated in a future version of the manuscript*

**SM7 - Hyperbolic heat equation solutions**

(Note: the equation numbers refer to each SM individual sub-section)

The 3D hyperbolic heat equation is solved using Finite Element Methods with COMSOL Multiphysics to calculate the phase lag between the harmonic laser excitation and the temperature response of the system. The laser energy deposition is restricted to a region defined by the Gaussian function of the pump beam in the radial direction and an exponential decay in the cross-plane direction with the characteristic length of the optical penetration $\Lambda_{\text{pump}} \equiv h = 14$ nm. The temperature oscillations correspond to a weighted average across the surface of the sample computed using the Gaussian function of the probe beam as the weight.

### A. Relation between the phase lag and the thermal penetration depth

The thermal penetration is defined as the depth at which the temperature increase due to the laser beam is attenuated by a factor 1/e with respect to the superficial temperature increase. In order to provide "a more intuitive description" of the frequency evolution of the phase lag, we provide the relation between the phase lag and the thermal penetration depth, which can be summarized as follows: **(i)** for the lower frequency regime, where heat conduction is dominated by diffusive heat transport, a larger phase lag ($\varphi$) implies smaller penetration depth, $\uparrow \varphi \implies \downarrow \Lambda_{diff}$, and **(ii)** for the higher frequency regime the penetration depth approaches $\Lambda_{ss} = 2\sqrt{\alpha \tau_{ss}}$, independently of the value of the phase lag, $\varphi$. The full interdependence between $\Lambda_{\text{HHE}}$ and $\varphi$ is displayed in Figure **SM7-1**. The derivation of the expression for $\Lambda_{\text{HEE}}$ as a function of the heating frequency is provided in section SM7C.

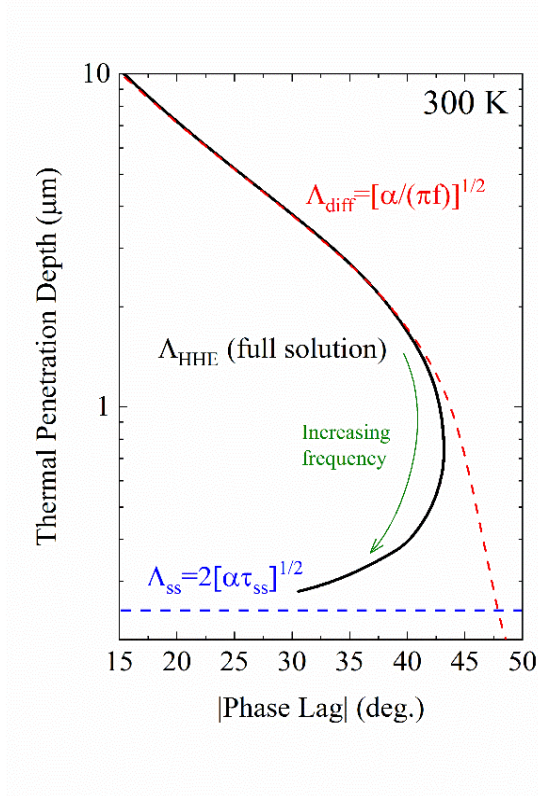

**Figure SM7-1**
Thermal Penetration depth calculated using the hyperbolic heat equation (HHE) as a function of the phase lag between the thermal excitation and the system response ($\Lambda_{HHE}$). The thermal penetration depth is also shown for the case of diffusive heat transport ($\Lambda_{diff}$), as well for the wave-like limit ($\Lambda_{ss}$). The green arrow indicates the direction of increasing excitation frequency.

### B. Influence of the pump and probe penetration depth

In this subsection we study the effect of increasing the penetration depth of the pump and probe lasers on the phase lag response. We show that if hypothetically considering a "larger effective pump penetration depth", the phase lag curves would exhibit a correspondingly larger Fourier behavior in thermal transport, i.e. the phase lag approaches $-\pi/4$ and even lower values (also shown through Eq. 15 of section SM7C). In other words, a larger effective pump penetration depth would mask, and not fictitiously amplify a second sound signature. Indeed, we show that the observation of second sound is not evident when the penetration depth of the heating region exceeds ≈100 nm.

We have performed finite element calculations using the HHE varying the penetration depth of the "heated region" (pump) and considering as superficial the detection region (probe), as well as varying the penetration depth of the detection region (probe) and considering a 14 nm heating region (pump). Note that the pump region cannot be set to be too superficial due to the energy conservation considerations. Figure **SM7-2** displays the results from these calculations. As the penetration depth of the pump increases, the phase lag gradually decreases even beyond $-\pi/4$.

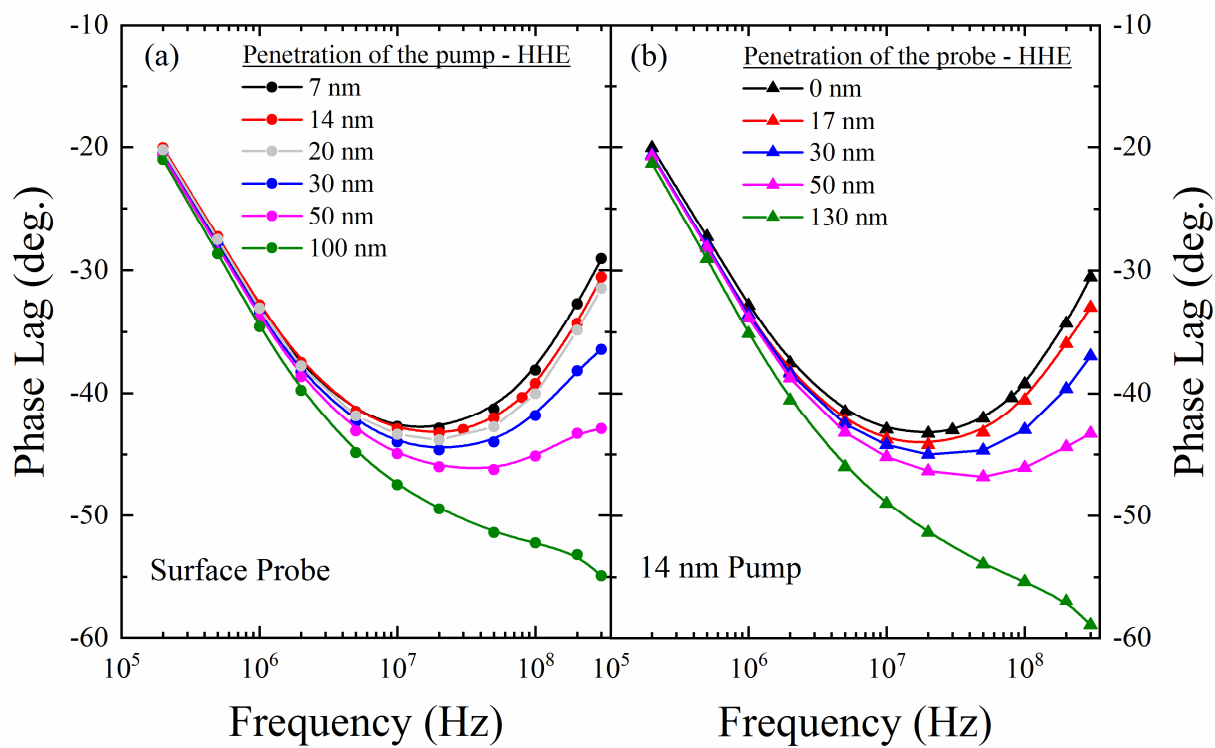

**Figure SM7-2**
Phase lag response vs excitation frequency simulated using the hyperbolic heat equation (HHE). (a) Influence of different penetration depths of the heating region in the phase lag response studied for the case of surface detection. (b) Influence of increasing the probe region on the phase lag response for heat source with 14 nm penetration depth.

C. **1D limit of the HHE**

As shown in Figure **1D**, the thermal penetration in the sample is reduced by increasing the excitation frequency. At room temperature and the highest experimentally available frequencies, the thermal penetration is significantly smaller than the pump laser spot radius. Therefore, thermal transport is restricted to the cross-plane direction and the system thermal response is 1D. Exploiting this, in this section we derive the analytical 1D hyperbolic equation solutions and hence we provide a simple expression for the phase lag at high frequencies and temperatures.

In the present experiment, the energy released to the semiconductor is introduced as a heat source with characteristic length $\Lambda_{pump} \equiv h = 14$ nm. Consider the energy conservation and the Maxwell-Cattaneo heat equations in 1D:

$$C_v \frac{\partial T}{\partial t} + \frac{\partial q}{\partial x} = \Sigma(x,t), \qquad (1)$$

$$q + \tau_{SS} \frac{\partial q}{\partial t} + \kappa \nabla T = 0 \qquad (2)$$

where $C_v$ is the specific heat, $\kappa$ is the thermal conductivity, $\tau_{SS}$ is the heat flux relaxation time and

$$\Sigma(x,t) = \frac{\Sigma_0}{h} e^{-\frac{x}{h} iwt} \qquad (3)$$

is the energy heat source, with $\Sigma_0 = 1$ W and $w$ is the laser heating angular frequency.

We look for stationary solutions of the form

$$C_v T(x,t) = F(x) e^{iwt}, \qquad (4)$$

$$q(x,t) = G(x) e^{iwt} \qquad (5)$$

According to equations (1,2), the functions F(x) and G(x) satisfy

$$iwF + G' = \frac{\Sigma_0}{h}e^{-\frac{x}{h}}, \qquad (6)$$

$$(1 + iw\tau_{SS})G = -\alpha F' \qquad (7)$$

where $\alpha$ is the thermal diffusivity. By defining

$$\gamma^2(w) = \frac{iw - \tau_{SS}w^2}{\alpha} \qquad (8)$$

and combining equations (6,7) we obtain the following second order nonhomogeneous differential equation:

$$\gamma^2 G - G'' = \frac{\Sigma_0}{h^2}e^{-\frac{x}{h}}. \qquad (9)$$

The general solution of (9) is the combination of the homogenous solution with negative exponent (heat flux vanish far away from the semiconductor surface at x=0) and a particular solution

$$G(x) = Ae^{-\gamma x} + G_0 e^{-\frac{x}{h}}, \qquad (10)$$

where $G_0 = \frac{\Sigma_0}{(\gamma h)^2 - 1}$ and A is a constant depending on the boundary conditions. We impose the insulation boundary condition $0 = q(x = 0, t) = G(x = 0) = A + G_0$ and we obtain

$$G(x) = G_0(e^{-\frac{x}{h}} - e^{-\gamma x}) \qquad (11)$$

Therefore, from (6),

$$F(x) = \frac{1}{iw}\left(G_0(\frac{e^{-\frac{x}{h}}}{h} - \gamma e^{-\gamma x}) + \frac{\Sigma_0}{h}e^{-\frac{x}{h}}\right) \qquad (12)$$

and the solution for the temperature reads

$$C_v T(x,t) = \frac{\Sigma_0}{iw}\left((\frac{G_0}{\Sigma_0}+1)\frac{e^{-\frac{x}{h}}}{h} - \gamma\frac{G_0}{\Sigma_0}e^{-\gamma x}\right)e^{iwt} \qquad (13)$$

The thermal penetration depth of the perturbation is then $\Lambda_{HHE} = 1/\Re(\gamma)$. At low frequencies $w\tau_{SS} \ll 1$, we recover the classical penetration depth $\Lambda_{diff} = \sqrt{2\alpha/w}$. By increasing the frequency, the penetration depth deviates from the classical prediction and, in the limit $w\tau_{SS} \gg 1$, it becomes frequency-independent: $\Lambda_{SS} = 2\sqrt{\alpha\tau_{SS}}$. Moreover, the wavelength of the thermal oscillations is $\lambda_{HHE} = 2\pi/Im(\gamma)$ (note that at high frequencies we obtain the limit $\lambda_{SS} = v_{SS}/f$ as expected). These characteristic lengths obtained from the 1D-HHE properly characterize the full 3D problem as confirmed by the Finite Elements calculations.

At the surface

$$C_v T_s = C_v T(x=0,t) = \frac{\Sigma_0}{iwh}\left(\frac{G_0}{\Sigma_0}(1-\gamma h) + 1\right)e^{iwt}. \qquad (14)$$

The resulting phase lag between the laser and the temperature oscillation is

$$arg\,[T_s] = -\frac{\pi}{2} + arctan\left(\frac{\sin\theta}{\cos\theta + h|\gamma|}\right), \quad (15)$$

where $\theta = arg\,[\gamma]$. From (15), it can be seen that by increasing the heating frequency to values close to $\tau_{SS}^{-1}$, the wave-like thermal behavior emerges leading to a non-monotonous behavior of the phase lag. In Figure **SM7-3** we show the 1D HHE solution and the 1D Fourier solution ($\tau_{SS}=0$) at room temperature compared with the corresponding 3D solutions from both models using Finite Elements calculations. It can be seen that the 1D solutions match the 3D numerical solutions for $f$ > 30 MHz. In fact, in this frequency range and at room temperature the thermal penetration $\Lambda_{HHE}$ is one order of magnitude smaller than the laser spot diameter and, thus, the system response is effectively restricted to 1D as expected. Conversely, at lower temperatures the thermal penetration depths are larger and the transition to the 1D behavior is not obtained within the studied frequency range. This allows to confirm the numerical fits of $\tau_{SS}$ at high temperatures using the analytical 1D-HHE solution as shown in Table **SM2**

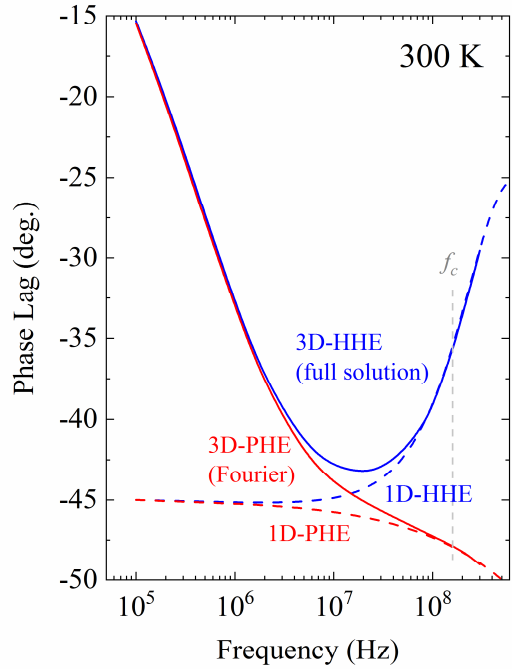

**Figure SM7-3**

Phase lag curves calculated using different heat transport models: (i) 3-dimensional hyperbolic heat equation (3D-HHE), (ii) 3-dimensional parabolic heat equation (3D-PHE, diffusive), (iii) 1-dimensional hyperbolic heat equation (1D-HHE), and (iv) 1-dimensional parabolic heat equation (1D-PHE, diffusive)

### D. Origin of the minimum of the phase lag curves and critical frequency ($f_c$)

In this subsection we address the origin of the minimum of the phase lag response curves and we show that, in general, the frequency of the minimum is not equivalent to $f_c$. As temperature decreases, $f_c$ gradually approaches the minimum of the phase lag curve as observed in Figure **SM2-2** and Figure **1D**. However, at room temperature $f_c$ is not well represented by the minimum. The origin of this effect is that at room temperature the system experiences a 3D→1D heat flow transition as frequency increases. Therefore, the reduction of the thermal dimensionality causes a shift of the phase lag minimum curve to frequencies smaller than $f_c$. Note that this effect is only observed at high temperatures in the present experimental conditions since, for higher frequencies ($f > 10$ MHz), the penetration depth ($\Lambda_{HHE} < 1$ μm) is much smaller than the diameter of the spot (≈11 μm) and, thus, heat transport become 1D. In contrast, this effect is gradually suppressed as temperature decreases since the thermal penetration depth increases, thus, leading to 3D heat transport within all the considered frequency range. In such conditions, the minimum of the phase lag curve is uniquely related to the unlock of wave-like effects, hence $f_c$ and the frequency of the minimum coincide. In Figure **SM7-4** we show calculations of the phase lag using the HHE in 3D and its 1D limit, and for different spot sizes of the heating region at room temperature. At high frequencies the 3D and 1D solutions of the HHE are equivalent, thus, showing that the system undergoes the mentioned 3D→1D transition. Consistently, at low frequencies the 1D limit of the HHE is not a good approximation to our experiments due to the finite size of the heating spot. However, as the spot size increases the solution at lower frequencies gradually approaches -π/4.

The frequency position of the minimum is dependent on three parameters: the thermal diffusivity of the sample ($\alpha$), the relaxation time of second sound ($\tau_{ss}$), and the size of the heating spot ($R_{spot}$). Figure **SM7-4** displays simulations for different diameters of the heat source ($2R_{spot}$), and using the room temperature values of $\alpha = 3 \times 10^{-5}$ m$^2$/s and $\tau_{ss} = 500$ ps. As discussed above, by reducing the spot size the 3D→1D transition is shifted to higher frequencies and, hence, the position of the minimum becomes closer to $f_c$. On the other hand, the influence of $\alpha$ and $\tau_{ss}$ on the position of the minimum is rather easier to qualitatively address. Exploiting the 3D →1D transition, we explain in a simple fashion how the minimum depends on $\alpha$ and $\tau_{ss}$ using the 3D-PHE and the 1D-HHE for the lower and higher frequency range, respectively. Figure **SM7-4** displays calculations of the 3D diffusive response (3D-PHE) for different values of $\alpha$. In addition, we also show the 1D limit of the HHE for different values of $\tau_{ss}$. All calculations were done for a fixed spot size, $2R_{spot}=10$ μm. The influence of $\alpha$ and $\tau_{ss}$ on the position of the minimum of the phase lag curve can be approximated by the intersection between the 3D-PHE and the 1D-HHE. Larger values of $\alpha$ imply larger values of the thermal penetration depth, which cause a shift to higher frequencies of the 3D→1D transition and, accordingly, the minimum of the phase lag curve is obtained at higher frequencies. Conversely, by increasing the value of $\tau_{ss}$ the wave-like effects (i.e. a reduction of the phase lag) are obtained at lower frequencies, which shift the minimum to lower frequencies.

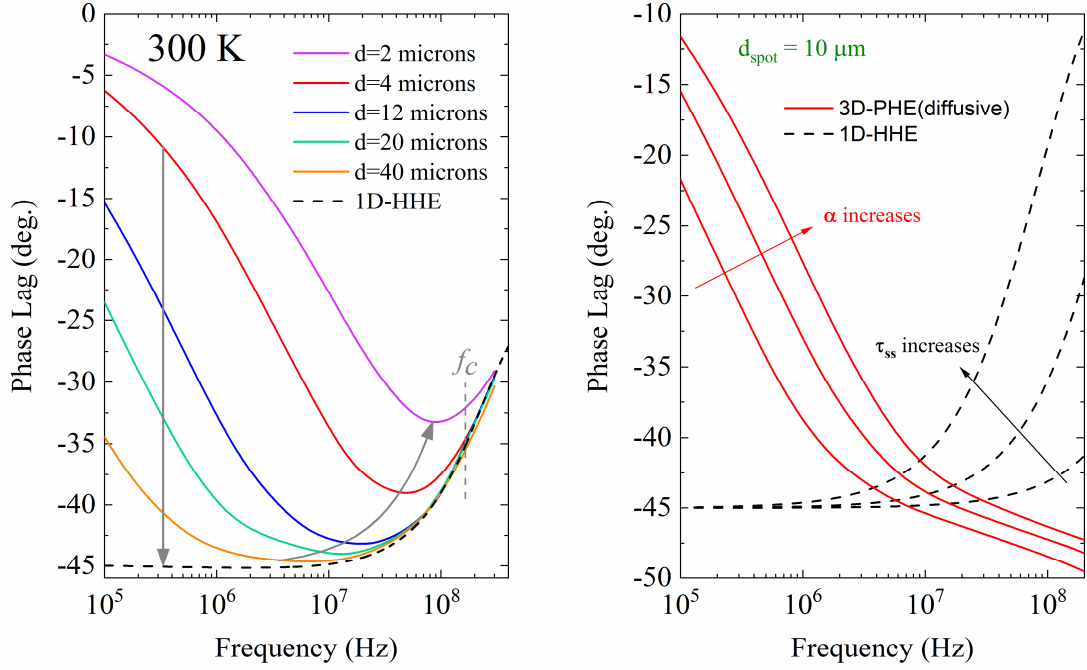

**Figure SM7-4**
(Left) Influence of different spot sizes diameter on the position of the minimum of the phase lag curves for $\alpha = 3x10^{-5}$ m²/s, and $\tau_{ss} = 500$ ps. The dashed line corresponds to the solution of the 1D-HHE (Right) Influence of increasing the thermal diffusivity ($\alpha$) and relaxation time ($\tau_{ss}$) on the minimum of the phase lag curve for d=11 μm. The approximate position of the minimum is given by the intersection of the 3D-PHE with the 1D-HHE.

### E. Temperature field evolution at the critical frequency: $f_c$

It is also interesting to address the thermal response of the system at $f_c$, since this is one of the key parameters describing the behavior of the system. As obtained in section SM7C, the thermal penetration depth at $f_c$ is $\Lambda_{HHE}(f_c) \approx 2.6\sqrt{\alpha\tau_{ss}}$. On the other hand, the wavelength of the second sound waves at the critical frequency is $\lambda_{HHE}(f_c) \approx 10\sqrt{\alpha\tau_{ss}}$ (see SM7C). Hence, the relation between the thermal wavelength and the thermal penetration depth is as follows:

$$\frac{\lambda_{HHE}(f_c)}{\Lambda_{HHE}(f_c)} \approx 4$$

The previous expression shows that at the critical frequency, $f_c$, the thermal wavelength is larger by a factor 4 compared to the thermal penetration depth. In consequence, the temperature oscillations of the temperature field are only insinuated at $f_c$.

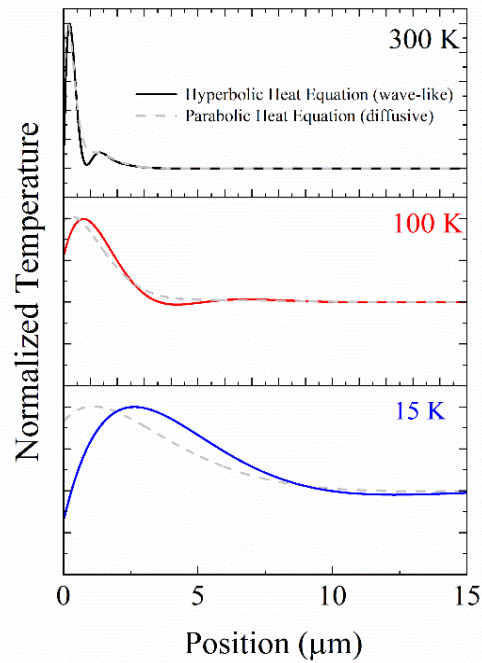

**Figure SM7-5**
Finite element simulations of the spatial distribution of the temperature field as a function of temperature in the direction perpendicular to the surface of the sample at the critical frequency $f_c$ at each temperature, and for an arbitrary time. The parabolic (diffusive) and hyperbolic solutions are shown in dashed and full lines, respectively.

However, the absence of temperature oscillations at $f_c$ does not imply that the thermal response is purely diffusive, but it rather means that the temperature response of the system is of mixed character, i.e. diffusive + wave-like. The spatial distribution of the temperature field at $f_c$ is shown in Figure **SM7-5**. Interestingly, although no oscillations are observed, the response of the system is different from the purely diffusive response, confirming the observation of partial wave-like behavior at $f_c$.

The relation between the wavelength of second sound and the dimensions of the system is also interesting to address. For example, considering the possibility of studying wave-like effects in polycrystalline samples. The observation of second sound in such conditions will largely depend on the size of the nano/micro crystals of the studied samples. We note that in order to observe second sound the size of the crystals must be larger than the thermal wavelength of the second sound wave. For example, in Ge at room temperature and at the highest frequency of $f = 200$ MHz, the wavelength of second sound wave is $\lambda = v_{ss}/f \approx 830\ nm$. Thus, it is expected that a nanocrystalline sample should not clearly exhibit wave-like effects since the size of the nanoscrystals will possibly limit the existence of second sound. In addition, it should be noted that in nanocrystalline sample the thermal conductivity is reduced by almost two orders of magnitude with the respect to the bulk case. The origin of this effect is, in part, a reduction of the phonon lifetimes. Hence, this lifetime reduction will also affect the relaxation of second sound, which is obtained through Eq. (2) in of the main text.

**SM8** – Local equilibrium conditions and definition of local temperature

The concept of temperature is well defined in equilibrium thermodynamics. However, its definition far from equilibrium is a fundamental issue of non-equilibrium statistical physics still under debate. Different temperatures can be defined based on different frameworks, such as macroscopic considerations, Fluctuation-dissipation theorems or different degrees of freedom in kinetic theory, and can provide different values (46–48). Below we discuss that, at the length and time scales of our experiments, we are mostly close to local-equilibrium, where all the definitions of temperature yield the same result. In addition, we also study quantitatively the applicability of the model developed within section SM6, demonstrating that this approach is expected to work well for T > 50 K within our experimental conditions.

### A. Local equilibrium

Temperature and thermal energy are quantities related to the system disordered energy. Under a heat flux, $q$, thermal energy flows more in one direction than in the opposite, thus, introducing some order. When the heat flow is small, most of the energy keeps disordered so that temperature and all thermodynamic properties can be defined locally and behave as in equilibrium (or local equilibrium). One can quantify how small is the heat flux by comparison with the limit situation where all the thermal energy flows in one direction, i.e. for a maximum heat flux $Q = c_v T v$, where $c_v T$ is the thermal energy and $v$ is the carrier velocity (46). Since the latter situation is far from equilibrium conditions, all energy becomes ordered (all carriers flow coherently) so that temperature has no meaning. In the opposite limit, $q \ll Q$, one is in local equilibrium. By introducing $q = -\kappa \nabla T$ this limit can be written as:

$$\kappa \nabla T \ll c_v v T \quad (1)$$

This allows to define a parameter characterizing the deviation from equilibrium (47),

$$a = \frac{q}{Q} = \frac{\kappa \nabla T}{c_v v T} = \frac{v \tau_k}{3L} \frac{\Delta T}{T} \quad (2)$$

where $\tau_k = \frac{3\kappa}{C_v v^2}$ is the mean free time ($\tau_k \approx 30$ ps in germanium at 300 K) and we have approximated $\nabla T \approx \Delta T/L$ with L the characteristic length scale for the temperature gradient. This parameter is analogous to $l \frac{\nabla T}{T}$, namely, the temperature variation ($\nabla T$) in a mean free path ($l \approx v \tau_k$) as compared to $T$, which must be small to assume local equilibrium (49). Local equilibrium thus requires:

$$a \ll 1 \quad (3)$$

### B. Memory Effects

For a complete description, memory effect should also be considered since they have direct implications on the validity of the definition of local temperature. In our case, memory is described through the hyperbolic equation:

$$\tau_{ss} \frac{\partial \vec{q}}{\partial t} + \vec{q} = -\kappa \overrightarrow{\nabla T} \quad (4)$$

Using Fourier transform and plane waves of the form $\exp[i(\omega t + \kappa x)]$, it is possible to obtain an algebraic relation between the relevant magnitudes related to our experiment,

$$(1 + \omega \tau_{ss}) \tilde{q} = -\kappa \widetilde{\nabla T} \quad (5)$$

Where $\tilde{q}$ and $\widetilde{\nabla T}$ are the Fourier transform of the heat and the temperature gradient, respectively. In this relation, the right side describes the excitation, and the left side the response of the system. The response will be dominated by the memory effects at frequencies $f > (2\pi\tau_{ss})^{-1}$. This frequency is (aside a factor 2) the characteristic frequency $f_c$ obtained by equating diffusive and wave-like penetration lengths. Let us recall that wave-like effects are noticeable for frequencies an order of magnitude lower, $f_c/10$, as shown in Figs. **2D** and **SM2-2**. Note that the relaxation time $\tau_{ss}$ appearing in equation (4) is different from $\tau_\kappa$ appearing in the thermal conductivity in the previous section. For Germanium at 300K, we have $\tau_{ss}$ = 500 ps vs $\tau_\kappa$ = 32 ps.

### C. Combination of local equilibrium and memory conditions

In the case that we excite the sample with a fast oscillation, when memory effects are important, the thermal excitation will propagate as a wave with a wavelength $L = v_{ss}/f$. Using this expression in Eq. (2) we obtain the condition of local equilibrium:

$$a = \frac{v}{3v_{ss}} f \tau_\kappa \frac{\Delta T}{T} \ll 1 \qquad (6)$$

Assuming $\Delta T \approx 5$ K, for the maximum frequency studied in the paper, $3\times10^8$ Hz, Eq. (6) yields $a \cong 3.65\times10^{-4}$ at 300 K. At 15 K, it provides $a \cong 8.76\times10^{-2}$. Hence, except below 15 K, the experimental range studied in the paper seems within the local equilibrium hypothesis.

### D. Non-equilibrium distribution function

The ab initio calculations of the relaxation time in the hyperbolic heat equation, Eq. (4), are based on a non-equilibrium distribution for the phonon population of the type:

$$f_\lambda = f_\lambda^{eq} + \vec{\beta}_\lambda \cdot \vec{q} + \vec{\gamma}_\lambda \cdot \frac{\partial \vec{q}}{\partial t} \qquad (7)$$

which is a perturbation from the local equilibrium distribution, $f_\lambda^{eq}$. It is expected that Eq. (7) is valid if the perturbation is small as compared to $f_\lambda^{eq}$. In the RTA approximation, where we have explicit expressions for coefficients $\vec{\beta}_\lambda$ and $\vec{\gamma}_\lambda$ (Eqs. (17) and (18) in **SM6**):

$$\vec{\beta}_\lambda = \frac{1}{\kappa} \frac{\partial f_\lambda^{eq}}{\partial T} \vec{v}_\lambda \tau_\lambda \qquad (8)$$

$$\vec{\gamma}_\lambda = \vec{\beta}_\lambda (\tau_\lambda - \tau_{SS}) \qquad (9)$$

it is possible to evaluate the relative size of the perturbations to $f_\lambda^{eq}$. For the first term in the right-hand side of (7) we obtain:

$$\frac{1}{\kappa} \frac{\partial \ln(f_\lambda^{eq})}{\partial T} \vec{v}_\lambda \tau_\lambda \cdot \vec{q} \approx \frac{q}{c_v T v} = a \ll 1, \qquad (10)$$

here we have approximated $v_\lambda \approx v$, $\tau_\lambda \approx \tau_\kappa$, and $\frac{\partial \ln(f_\lambda^{eq})}{\partial T} \approx \frac{1}{T}$. Note that the inequality in Eq. (10) arises from the condition of local equilibrium discussed above.

For the second term of Eq. (7) we obtain:

$$\frac{1}{\kappa} \frac{\partial \ln(f_\lambda^{eq})}{\partial T} \vec{v}_\lambda \tau_\lambda (\tau_\lambda - \tau_{SS}) \omega q \approx a\omega\tau_{SS} \equiv b \qquad (11)$$

Therefore, the perturbative expansion requires both $a \ll 1$ and $b \equiv a\omega\tau_{SS} \ll 1$. Assuming again $\Delta T \approx 5K$ and the maximum frequency studied of $3\times10^8$ Hz, one obtains $b \cong 3\times10^{-4}$ at 300 K, and $10^{-1}$ at 50 K. At 15 K we obtain $\approx 1$, though the RTA approximation is not a good approximation at such low temperatures. However, the trend observed as temperature decreases suggests that b could reach values higher than 1. As a result of the previous calculations, above 50 K we thus expect the ansatz Eq. (7) to be a reasonable approximation in the frequency range studied, but we may not expect it to be so good at lower temperatures. Indeed, Figs. **SM2-2** (d) and (e) show some deviations from the HHE.

Finally, we estimate the values for the *a* and *b* parameters as a function of temperature as follows:

| Temperature [K] | a | b |
|---|---|---|
| 7 | $3.6 \times 10^{-1}$ | 2.82 |
| 15 | $8.76 \times 10^{-2}$ | $7.59 \times 10^{-1}$ |
| 50 | $1.83 \times 10^{-2}$ | $1.27 \times 10^{-1}$ |
| 100 | $5.19 \times 10^{-3}$ | $2.05 \times 10^{-2}$ |
| 300 | $3.65 \times 10^{-4}$ | $3.44 \times 10^{-4}$ |

**Table SM8-1:** Parameters *a* and *b* defined through of Eq. (7) as a function of temperature.

### E. Local temperature estimation through Molecular Dynamics calculations.

We also aim to show to which extent temperature is well defined in our experiments through Molecular Dynamics calculations. For this purpose, according to the local equilibrium principle of non-equilibrium thermodynamics, we have calculated the "local and instantaneous" temperature of three regions in the simulated Ge sample, namely: (i) the true hot thermostat with varying temperature, (ii) a thin slab 10 nm away from the thermostat within the Ge sample (i.e. nearby the thermostat), and (iii) a thin slab 40 nm away from the thermostat within the Ge sample (i.e. at a long distance from the thermostat). By "local" it must be understood that the slabs were chosen so as to have a thickness (2.7 nm) much smaller than the total system size. By "instantaneous" it must be understood that the time-average was taken over a time lapse. much shorter than the period of time variation of the T(t) signal. The calculations have been done in the most unfavorable condition, i.e. with a very high frequency of temperature oscillation, actually 5 GHz. Temperature was calculated from the average kinetic energy of the atoms within the selected slab.

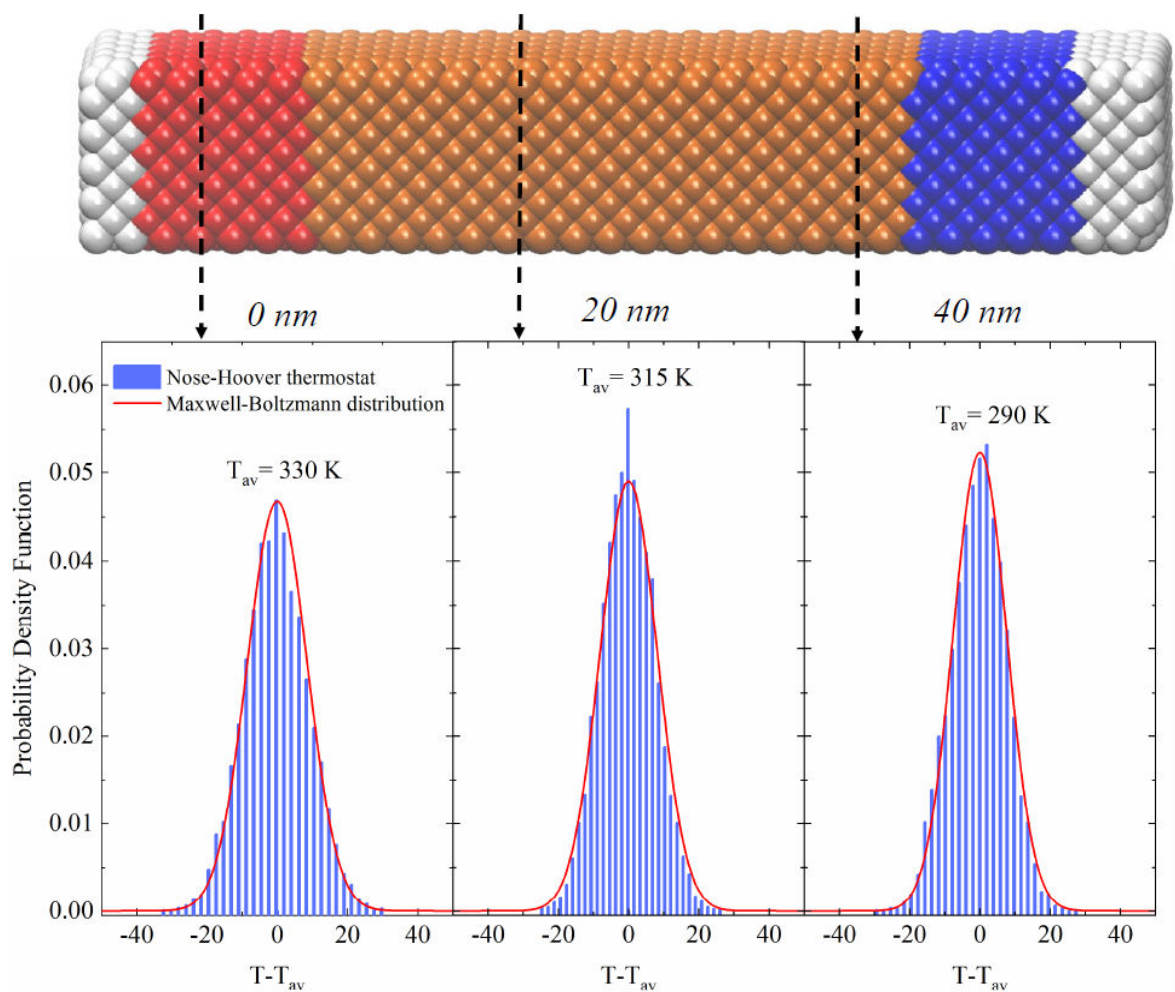

**Figure SM8-1:** Local temperature obtained through molecular dynamics calculations in different positions of the sample at 5 GHz. The red full lines are fit to the data points using a Maxwell-Boltzmann distribution.
.

The results are shown in Figure **SM8-1**, where it is very cleanly proved that the "local and instantaneous" temperature is distributed around its mean value (blue histograms) with a Maxwell-like distribution (red curve). Please note that the Maxwell law of distribution for the atomic velocities (or temperature, it is just a matter of normalizing factors) has been calculated at the mean temperature value: this is a robust choice, since the Maxwell law is only marginally affected by a temperature variation of +/-10-20 degree Kelvin, as in fact found out of the MD data.

**SM9** – Influence of the Au transducer on the high frequency response

In this section we study the effect of introducing a thermal interface through the deposition of a gold metallic transducer. This approach has been widely used to measure the thermal conductivity and heat capacity of substrates and thin films (50,51). Figure SM9-1 displays the frequency dependent phase lag for two different interfaces, (i) Ge/native-$GeO_2$/Au and (ii) Ge/Au. The data for bare Ge is shown for relative comparison. The thermal interface in (ii) is expected to be a better thermal conductor than (i) due to the lack of the oxide layer. In both cases, the observed response differs substantially from the results obtained for bare Ge. The presence of the thermal interface between the Au transducer and the Ge substrate, as well as the Au transducer itself, dominate the system response in the frequency range where wave-like effects are expected. In particular, this becomes the dominant contribution over 100 MHz, which we have verified studying different thicknesses of the Au transducer, as well as measuring the system response on different substrates. These observations strengthen the importance of having chosen Ge that, thanks to its absorption characteristics discussed above, is suited for thermoreflectance experiments even without a metallic transducer, which, as shown in Figure SM9-1, would hinder the observation of second sound.

The presence of the thermal interface between the Au transducer and the Ge substrate, as well as the Au transducer itself, dominate the system response in the frequency range where wave-like effects are expected. The combined effect of the transducer and of the thermal boundary resistance with the substrate in the high frequency range has been recently discussed in Ref. (51). Although this effect is interesting by itself, the main purpose of these measurements is to show that using a Au transducer do not lead to the observation of second sound.

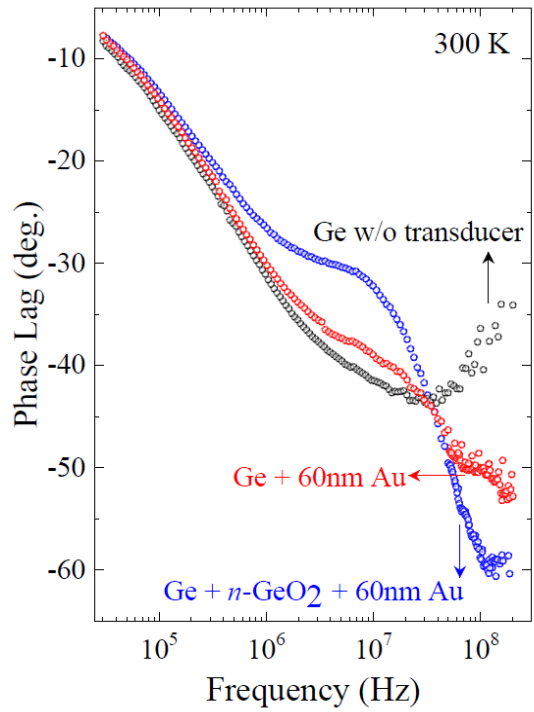

**Figure SM9-1**
(Influence of the presence of a thermal interface on the frequency-dependent phase lag at room temperature.

wavelength v_ss/f. At room temperature and the highest excitation frequency considered v_ss/f=840 nm, thus, considerably larger than the optical penetration depth.